\documentclass[twocolumn]{emulateapj}

\usepackage{amsmath,epstopdf,natbib,apjfonts,graphicx}

\newcommand{\bootes}{Bo\"{o}tes} 
\newcommand{\xbootes}{XBo\"{o}tes} 

\newcommand{\lagn}{L_{\rm AGN}}
\newcommand{\lsf}{L_{\rm IR}^{\rm SF}}
\newcommand{\ltot}{L_\textmd{IR}^\textmd{tot}}

\newcommand{\ergs}{$\textmd{erg}\;\textmd{s}^{-1}$}

\newcommand{\pz}{{\it photo-z}}
\newcommand{\pzs}{{\it photo-z}s}

\shortauthors{CHEN ET AL.}
\slugcomment{Accepted for publication in ApJ}

\begin{document}

\title{A Connection between Obscuration and Star Formation in Luminous Quasars}
\author{Chien-Ting J. Chen\altaffilmark{1}}
\author{Ryan C. Hickox\altaffilmark{1}}
\author{Stacey Alberts\altaffilmark{2,3}}
\author{Chris M. Harrison\altaffilmark{4}}
\author{David M. Alexander \altaffilmark{4}}
\author{Roberto Assef \altaffilmark{5}}
\author{Michael J.~I. Brown\altaffilmark{6}}
\author{Agnese Del Moro\altaffilmark{4}}
\author{William R. Forman \altaffilmark{7}}
\author{Varoujan Gorjian\altaffilmark{8}}
\author{Andrew D. Goulding\altaffilmark{7,9}}
\author{Kevin N. Hainline \altaffilmark{1}}
\author{Christine Jones \altaffilmark{7}}
\author{Christopher S. Kochanek \altaffilmark{10}}
\author{Stephen S. Murray\altaffilmark{11}}
\author{Alexandra Pope\altaffilmark{2}}
\author{Emmanouel Rovilos \altaffilmark{4}}
\author{Daniel Stern \altaffilmark{8}}

\altaffiltext{1}{Department of Physics and Astronomy, Dartmouth College, 6127 Wilder Laboratory, Hanover, NH 03755, USA; ctchen@dartmouth.edu.}
\altaffiltext{2}{Department of Astronomy, Amherst, University of
  Massachusetts, Amherst, MA 01003, USA} 
\altaffiltext{3}{Steward Observatory, University of Arizona, Tucson,
  AZ 85721, USA}
\altaffiltext{4}{Department of Physics, Durham University, South Road,
  Durham, DH1 3LE, United Kingdom}
\altaffiltext{5}{N\'ucleo de Astronom\'ia de la Facultad de
  Ingenier\'ia, Universidad Diego Portales, Av. Ej\'ercito Libertador
  441, Santiago, Chile}
\altaffiltext{6}{School of Physics, Monash University, Clayton 3800, Victoria, Australia.}
\altaffiltext{7}{Harvard-Smithsonian Center for Astrophysics, 60
  Garden Street, Cambridge, MA 02138.}
\altaffiltext{8}{Jet Propulsion Laboratory, California Institute of
  Technology, 4800 Oak Grove Dr., Pasadena, CA 91109, USA}
\altaffiltext{9}{Princeton University, Department of Astrophysical
  Sciences, Ivy Lane, Princeton, NJ 08544, USA}
\altaffiltext{10}{Department of Astronomy, Ohio State University, 140 West 18th Avenue, Columbus, OH 43210}
\altaffiltext{11}{Department of Physics \& Astronomy, The Johns Hopkins University, 3400 N.\ Charles Street, Baltimore, MD 21218.}

\begin{abstract}
We present a measurement of the star formation properties of a
uniform sample of mid-IR selected, optically unobscured and obscured quasars
(QSO1s and QSO2s) in the \bootes\ survey region. 
We use a spectral energy distribution (SED) analysis for photometric data spanning optical to far-IR wavelengths to
separate the AGN and host galaxy components. 
We find that when compared to a matched sample of QSO1s, the
QSO2s have roughly twice the higher far-IR detection fractions, far-IR fluxes and
infrared star formation luminosities ($L_{\rm IR}^{\rm
  SF}$). Correspondingly, we show that the AGN
obscured fraction rises from 0.3 to 0.7 between $(4-40)\times10^{11}L_\odot$. We also find evidence associating X-ray
absorption with the presence of far-IR emitting
dust. Overall, these results are consistent with galaxy evolution
models in which quasar obscuration is associated with a
dust-enshrouded starburst galaxies.
\end{abstract}


\section{Introduction}
Quasars, the most luminous active galactic nuclei (AGNs \footnote{In this work, we use AGNs as a general
  nomenclature for all energetically relevant SMBHs. For AGNs with bolometric
luminosity larger than $\sim10^{45}$ erg s$^{-1}$, we refer to them as
quasars (QSOs).}), have been linked to galaxies with
active star formation (SF) ever since the discovery of their observational
connection to ultra-luminous infrared galaxies \citep[ULIRGs, galaxies more
luminous than $10^{12} L_{\odot}$, e.g.][]{sand88}. 
One well-studied scenario for massive galaxy evolution posits that 
gas-rich galaxy major mergers trigger both rapid supermassive black
hole (SMBH) accretion and intense
SF. This scenario associates the dust-enshrouded starburst with
strong nuclear obscuration that is later expelled by the powerful AGN, implying an evolutionary link between
unobscured (type 1) and obscured (type 2) quasars
\citep[e.g.][]{dima05qso,hopk06apjs,gill07cxb,some08bhev,trei09}. 

On the other hand, the ``unification model'' of AGN ascribes obscuration of
AGNs to different lines of sight through a dusty ``torus'' surrounding the SMBH
\citep[e.g.][]{urry95std,anto93std}. This model predicts no difference in host galaxy properties between obscured and unobscured AGNs. 
To date, it is still a matter of debate whether the obscuration in
luminous quasars can be explained solely by the orientation-based
unification model or if it is also enhanced due to dust on larger
scales throughout the host galaxy. 
Several studies have shown results supporting a scenario departing
from the unification model, such as the enhanced SF activity
\citep[e.g.][]{cana01qso,page04submm,hine09,brus09highz,shan12} and
the more disturbed structure \citep[e.g.][]{lacy07qso2} of the host
galaxies of dust-obscured quasars when compared to unobscured
quasars. Clustering of different types of AGNs have also shown that
obscured AGNs are more strongly clustered than unobscured AGNs \citep[e.g.][]{hick11clust,dono13clust,dipo14}.
However, other studies have found no significant
difference between obscured and unobscured AGN populations in their
morphological and SF properties \citep[e.g.][]{stur06,zaka06host,zaka08,main11obsqso,scha12obsqso,merl13},
and the host galaxy star formation rate (SFR) does not distinguish X-ray selected AGNs with different
obscuring column densities \citep[e.g.][]{rovi12,rosa12agnsf,merl13}. \par

Nonetheless, the measurements of SF properties for quasar
host galaxies still suffer from selection biases that are often
different among various quasar populations. 
In particular, optical and X-ray selected quasar samples might have
different completeness in obscured and unobscured sources due to the
attenuation of optical and X-ray radiation by dust and gas. 
In fact, some studies have suggested that optical surveys might
miss $\sim 50\%$ of the AGN population due to both the obscuration and
host galaxy contamination in low-luminosity AGNs \citep[e.g.][]{goul09irs,goul10compthick}. 
While current X-ray observations probing photons with energy at $ \sim 10$ keV are believed to be less affected by moderate
levels of obscuring materials ($N_{\rm H} < 10^{24}$cm$^{-2}$ ), a
significant fraction of X-ray AGNs \citep[$\sim20-50\%$,
e.g.][]{donl05radio,guai05,park10,alex11stack,geor13,wilk13} might still be
missed due to obscuration that is Compton-thick ($N_{\rm H} \sim
10^{24}$cm$^{-2}$). Although very hard X-ray photons can penetrate Compton-thick
obscuration, the previous high energy X-ray surveys are
limited to local sources due to the shallow flux limits
(e.g. {\it Swift/BAT}, \citealt{burl11} and {\it INTEGRAL},
\citealt{sazo12}). While the {\it NuSTAR} \citep{nustar}
mission opened a new window of high energy X-ray up to 80 keV, the 
recent studies that used {\it NuSTAR} to observe heavily obscured AGNs
have found that obscuration is still a non-negligible effect even
for X-ray photons at such high energy \citep{ster14nustar,lans14nustar}.
\par

In contrast, mid-IR observations of the reprocessed emission from the
obscuring dust can detect heavily obscured AGNs. A number of studies have shown that large populations of AGNs can be
selected using the power-law SED shape of AGNs at mid-IR wavelengths \citep[e.g.][]{lacy04,ster05,hick07obsagn,donl12,asse13wise}.
Although these mid-IR color selection criteria cannot avoid some star forming
galaxy interlopers and might miss AGNs accreting at lower
accretion rates \citep[e.g.][]{hick09,donl12,mate13,hain14,chun14} or AGNs with
complicated silicate features in their mid-IR SED \citep{kirk13}, 
they are effective in selecting both obscured and
unobscured AGNs with similar completeness in mid-IR
wavelengths. Recent studies have shown that mid-IR selected AGNs can also be
separated into obscured and unobscured populations using a simple
optical to mid-IR color selection criterion
\citep[e.g.][]{hick07obsagn,dono13clust,dipo14}, which can easily be explained by the different level of
extinction in the optical emission from the nucleus \citep[e.g.][H07
hereafter]{hick07obsagn}. 
Therefore, a direct comparison of the host galaxy SF properties of obscured and unobscured sources in a mid-IR selected
quasar sample can provide insights to the origin of obscurations in rapidly accreting SMBHs. 

In this work, we adopt the mid-IR selected quasar sample from
\citet[H11 hereafter]{hick11clust}, which is comprised of unobscured and obscured quasars
with similar distributions in quasar properties (e.g. redshift and quasar luminosity), thus making it an excellent sample to study the 
connection between host galaxy properties and quasar obscuration.
The quasar sample studied in this work consists of 546 unobscured quasars (QSO1s) and 345 obscured quasars (QSO2s) selected using
{\it Spitzer} mid-IR observations in the \bootes\ survey region. We utilize
the optical spectroscopy from the AGN and Galaxy Evolution Survey 
\citep[AGES,][]{AGES} and the \xbootes\ {\it Chandra} X-ray observations \citep{murr05},
along with the $250\;\micron$ data from the Spectral
and Photometric Imaging Receiver \citep[SPIRE,][]{spire} on board the
{\it  Herschel Space Observatory}. 
With the inclusion of the far-IR photometry in which AGN contributions have
been shown to be much smaller than that of the host galaxy
\citep[e.g.][]{netz07qsosf,lacy07,kirk12,mull12agnsf,chen13sfagn,delm13,drou14}, 
we can use the wealth of multiwavelength observations in the \bootes\
field to obtain robust measurements of SFR for luminous quasars and
test if they are associated with obscuration.\par

We also take advantage of this sample to study the correlation
between star formation and AGN accretion for the mid-IR QSOs. 
In galaxy evolution models which suggest that the concurrent growth of AGN and host galaxy occurs during the dust
enshrouded galaxy merger phase, different types of quasars represent different
stages of galaxy evolution. To date, observational studies on the
connection between SFR and AGN accretion rate have not yet shown
definitive conclusions. Positive, strong correlations have been found in
studies of optically selected type I quasars \citep[e.g.][]{serj09qsosf} and type II quasars \citep[e.g.][]{netz09}; while different conclusions have been found in
X-ray selected quasars at higher redshift \citep[e.g.][]{silv09sfagn,mull12agnsf,rosa12agnsf}. 
The difference between observed correlations could be driven by various limitations and observational
constraints such as sample size \citep{page12,harr12} or AGN intrinsic
variability \citep[e.g.][]{chen13sfagn,hick14agnvar}. In addition, it is also important to note
that AGNs selected with different criteria are hosted by very different host galaxy populations
\citep[e.g.][]{hick09,grif10,goul13} in which the galaxy and SMBH might
follow different evolutionary paths. The sample in this work consists
of a large number of luminous QSOs including both the unobscured and
heavily obscured populations, thus biases due to small number
statistics and the exclusion of dust-enshrouded AGNs are small. 
The bias in the observed SFR-AGN accretion rate relation in AGN host
galaxies due to short-term AGN accretion rate stochasticity may also be less significant in luminous quasars \citep{hick14agnvar}.
Therefore, the SFR-AGN accretion rate correlation for the mid-IR selected
quasar populations may shed light on the origin of the concurrent growth of SMBH and galaxy. \par

This paper is organized as follows: in \S 2 we describe the
multi-wavelength data and the properties of the quasar sample. In \S4, we discuss the SED fitting procedures that we used to
disentangle the AGN and the host galaxy contributions.  In \S3,
we explore details of the observed far-IR properties of QSO1s and QSO2s. 
A comparison of the SF luminosity between QSO1s and QSO2s is laid out
at \S5; The x-ray properties for the QSO2s are discussed in \S6 and
the AGN obscured fraction as a function of SF luminosity is discussed
in \S7. A discussion and a summary are given in \S 8. Throughout the
paper, we use the Vega magnitude system and assume a $\Lambda$CDM cosmology with $\Omega_m=0.3$, $\Omega_\Lambda=0.7$ and $H_0=70$  km s$^{-1}$.

\begin{figure*}[t]
\epsscale{1.1} 
\plottwo{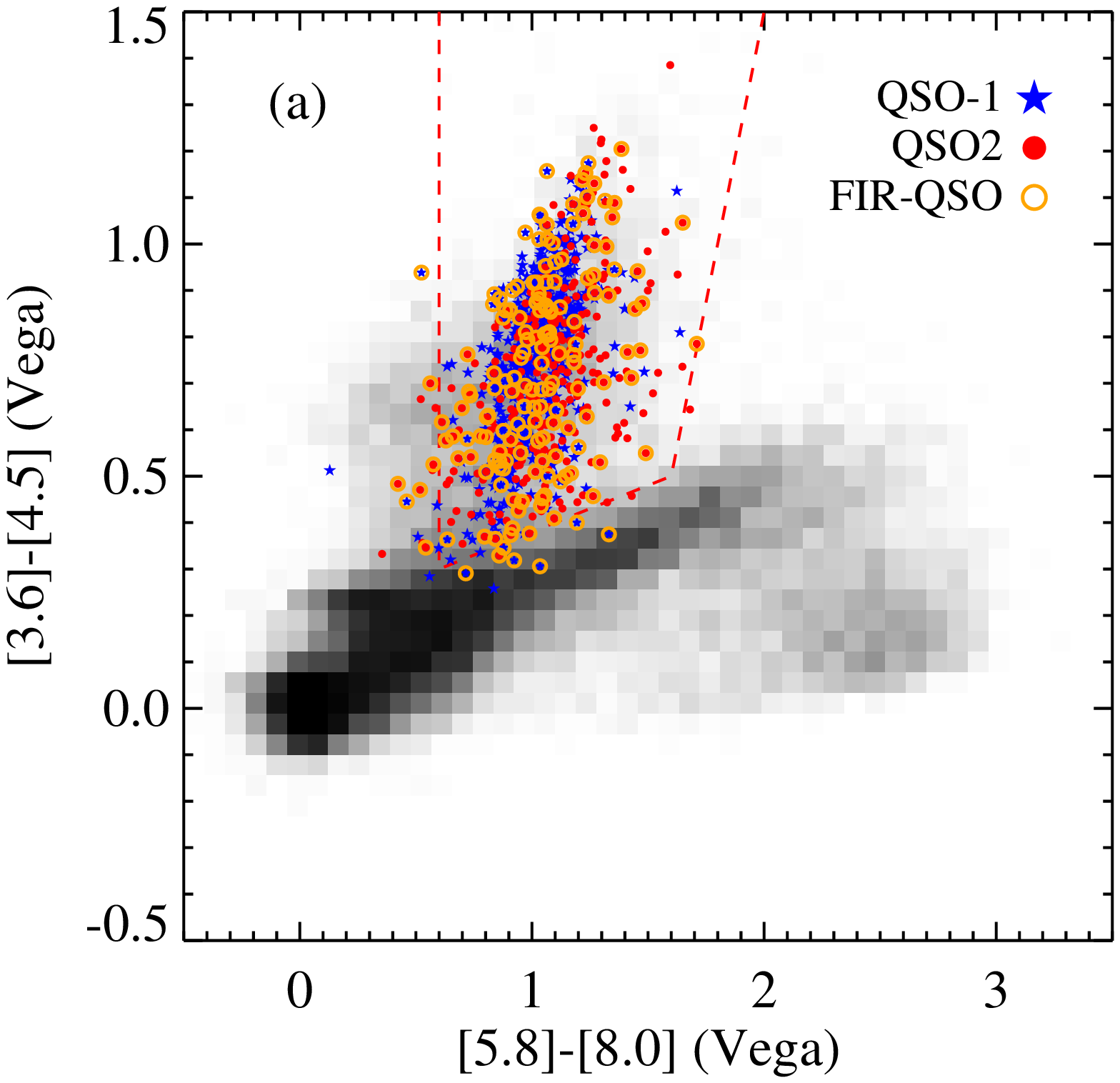}{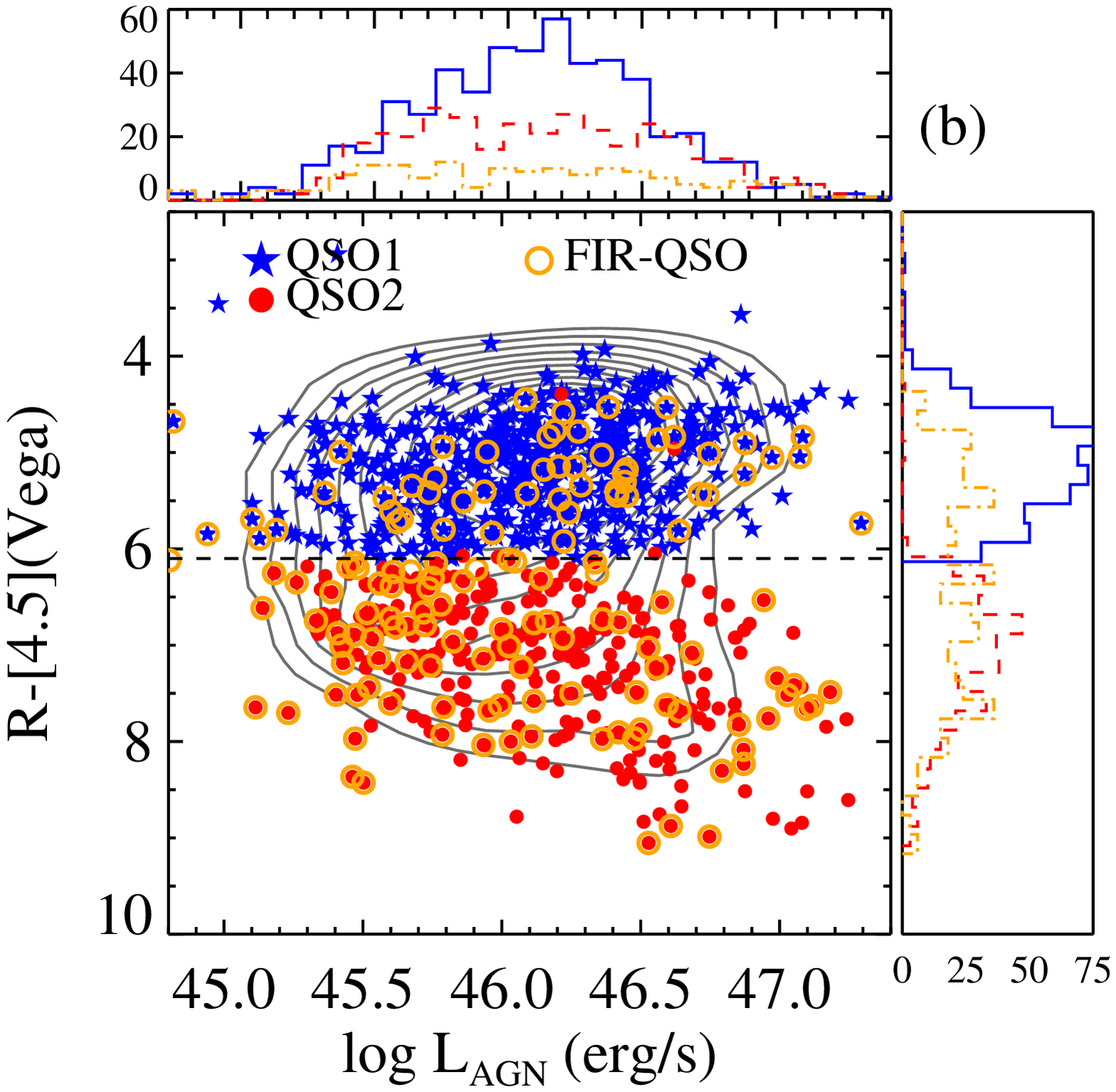}
\caption{(a) IRAC color-color diagram showing the selection of the
  quasar samples using the criteria of \citet{ster05}.  The gray-scale
  shows the density of sources detected at $>5\sigma$ significance in
  all four bands in SDWFS data.  Blue stars and red
  circles show the QSO1 and QSO2 samples, respectively.  The
  \citet{ster05} color-color selection region is shown by the dashed
  line. In addition, the QSOs with far-IR detections (FIR-QSO) are
  enclosed with orange circles. Some of the mid-IR QSOs fall out of the
  selection wedge due to the updated IRAC
  photometry and aperture corrections (see \S2.3).
(b) Illustration of the optical-IR
  color-selection criteria for dividing the IR-selected QSO sample
  into unobscured (QSO1) and obscured (QSO2) subsamples.
  Shown is observed ${\it R}-[4.5]$ color versus bolometric luminosity,
  calculated as described in \S 4.  Contours show the
  distribution for all the \cite{hick07obsagn} IR-selected quasars,
  while blue stars and red circles  show the QSO1 and QSO2
  subsamples at $0.7 < z < 1.8$ used in H11 and this analysis as described in
  \S 2. The distribution in the ${\it R}-[4.5]$ color and $L_{\rm AGN}$
  are also shown as histograms in the side panels. QSO1s are shown as the
  blue solid lines, QSO2s are shown as the red dashed lines and the
  FIR-QSOs are shown as the orange dash-dotted lines. The contours and color histograms
  show that a simple cut in optical-IR color clearly separates the
  QSO samples into two populations. }
\label{fig:sample}
\vspace{1.5ex}
\end{figure*}

\section{The Quasar Sample}
In this section, we discuss the data from the \bootes\ survey region as well as the quasar selection and classification criteria we adopted. \par

\subsection{Data}
The sample studied in this work comes from the 9 deg$^2$ \bootes\ survey region covered by the NOAO Deep Wide-Field Survey
\citep[NDWFS,][]{NDWFS}. \bootes\ is unique among extragalactic surveys because of its large area and the excellent
multi-wavelength coverage from space- and ground-based telescopes,
which make possible statistical study of the rare luminous AGNs. \par

In this work, we use the multiwavelength photometry catalog from Brown
et al. (private communication), which the same the one used in
\cite{chun14}. This catalog covers the optical to mid-IR bands including
the NDWFS optical observations in {\it Bw, R, I } bands, near-IR photometry from the NOAO NEWFIRM survey \citep[{\it
J, H} and {\it Ks},][]{gonz10} and the {\it Spitzer} Deep Wide Field
Survey \citep[SDWFS,][]{ashb09sdwfs} of the 4 bands of mid-IR
observations from the {\it Spitzer} Infrared Array Camera (IRAC) at
$3.6, 4.5, 5.8$ and $8\;\micron$.
In addition, the mid-IR photometry from the {\it Spitzer} Multi-band Imaging Photometer (MIPS)
at $24\;\micron$ is also included (IRS GTO team, J. Houck (PI), and M. Rieke). In this catalog, the optical to near-IR
photometry was extracted from a matched aperture for each band. For
{\it Bw, R, I, H} and {\it Ks} bands, the photometry was measured from images smoothed to a common
point spread function (PSF) with a 1\farcs35 full width half maximum
(FWHM); while for {\it J} band, the photometry was measured from
images smoothed to a common PSF with a 1\farcs60 FWHM. 
To ensure consistency in the photometry, we use 6\arcsec aperture
photometry from optical bands through the IRAC bands in the mid-IR to
  account for the large IRAC beam size. For the MIPS data, we use PSF
  photometry instead of aperture photometry due to the still larger beam size of MIPS. The $5\sigma$ flux limits of the optical to near-IR broad-band photometry are
25.2, 23.9, 22.9, 21.1, 20.1, 18.9 (Vega magnitudes) for the {\it Bw, R, I J, H, Ks}
bands, respectively. For the mid-IR wavelengths, the $5\sigma$ flux
limits are 6.4, 8.8, 51, 50 and 170 $\mu$Jy for the $3.6,4.5,5.8, 8.0$
and $24\;\micron$ bands, respectively. A extensive description of
the multiband photometry extraction can be found in \cite{brow07red}.\par

We also make use of the far-IR observations from the publicly
available {\it Herschel} Multi-tiered Extragalactic Survey
\citep[HerMES,][]{hermes}. We re-reduced and mosaiced the \bootes\
SPIRE observations \citep{albe13stack}, which include a
deep $\sim2$ deg$^2$ inner region near the center of the field and a
shallower $\sim8.5$ deg$^2$ outer region. We specifically focused on
removing striping, astrometry offsets, and glitches missed by the
standard pipeline reduction. We also convolved the raw maps with a
matched filter \citep[see][]{chap11}, which aided in source extraction
by lowering the overall noise and de-blending sources. From this, we
generated a matched filter catalog with a $5\sigma$ detection threshold.
In this catalog, we consider SPIRE sources with fluxes larger than
$20$ mJy as unambiguously detected. 
Completeness simulations show that these catalogs are $95\%$ complete
in the inner region and $69\%$ complete in the outer regions above a
flux of 20 mJy. We also find minimal flux boosting for low SNR
sources above this flux cutoff \citep[see S2.2 in][for the details
of the completeness simulation]{albe13stack}.
We match the positions of the
SPIRE catalog to the {\it I}-band positions with a matching radius
of 5\arcsec. We tested the rate of spurious matches by offsetting the
SPIRE source positions by $1\arcmin$ in a random direction and
matching the randomly shifted catalog to the \bootes\ catalog.
We found that with a radius of $5\arcsec$, our matching between the
SPIRE and \bootes\ catalog only yielded $<2\%$ spurious matches. 
\par

In addition, we also match the positions of the \cite{brow07red} catalog
to the publicly available  {\it Wide-field Infrared Survey Explorer}
(WISE) All-sky catalog \citep{wisecatalog} and obtain the profile-fit
photometry magnitudes in the W1, W2, W3 and W4
bands ($3.4, 4.6, 12$ and $22\;\micron$). \par

As a complementary measurement of the AGN accretion rate and
absorption by gas, we utilize X-ray data from the  XBo\"{o}tes survey, which is a mosaic of
126 short (5 ks) {\it Chandra} ACIS-I images \citep{murr05,kent05}
covering the entire NDWFS. XBo\"{o}tes contains 3,293 X-ray point
  sources with at least four counts in the AGES survey region. The
  conversion factors from count rates (in counts s$^{-1}$) to flux (in
  \ergs\ ) for the XBo\"{o}tes are $6.0 \times 10^{−12}$ \ergs\
  count$^{-1}$ in the 0.5--2 keV band and $1.9\times 10^{−11}$ \ergs\
  count$^{-1}$ in the 2--7 keV band, which are derived for a 5ks
  on-axis observation and assuming a canonical unabsorbed AGN X-ray
  spectrum \citep[see \S3.3 of][]{kent05}. \par

For this study, we use spectroscopic redshifts ({\it spec-z}s)  from AGES when possible. For the sources without a
{\it spec-z}, we adopt the photometric redshifts ({\it
  photo-z}s) calculated using techniques combining 
artificial neural network and template fitting algorithms
\citep{brod06photoz}. The uncertainty of this set of \pz\ is 
$\sigma=0.06(1+z)$ for galaxies and $\sigma=0.12(1+z) $ for AGNs. 
We note that the uncertainty for AGNs are dominated by the difference
between \pzs\ and {\it spec-z}s for the type I AGNs.
The uncertainties in the \pzs\ for QSO2s
are difficult to estimate accurately as only $9\%$ of the QSO2s have {\it
  spec-z} measurements. For these 32 QSO2s, the
uncertainty of the \pzs\ is
$\sigma=0.06(1+z)$, which is consistent with the uncertainty for the
galaxy population. Recently spectroscopic follow-up of 35
{\it WISE}-selected, optically obscured quasars by \cite{hain14} also
found that \pzs\ derived from template-based algorithm \citep{asse10sed}
have a similar accuracy. This is not surprising since the optical SED for the heavily obscured AGNs is dominated by the
host galaxy which has more spectral features and should allow for a more
accurate \pz\ measurement. \par

An upper limit for the QSO2 \pz\ uncertainties can be estimated based
on different approaches. H07 obtained an uncertainty of 
$\sigma_{z} = 0.25(1+z)$ by comparing the \cite{brod06photoz} \pzs\ to the
\pzs\ estimated by fitting 3 different galaxy templates to the {\it Bw, R, I}
photometry. Since the accuracy of the \pzs\ based on fitting 3 galaxy template
to 3 optical photometric observations is limited, the $\sigma_{z}
= 0.25(1+z)$ uncertainty is a very conservative upper limit. 
For this study, we adopt the \pz\ uncertainty upper
limit of $\sigma_{z} = 0.25(1+z)$ from H07. However, this is a very
conservative estimate since the \pz\ uncertainties for the QSO2s with
host galaxy dominated optical SEDs are likely to be much smaller. A
full discussion of the \pz\ uncertainties is given in \S8.

\subsection{AGN identification and classification}
The AGNs in this work is drawn from the quasar sample of H11, which is
a subset of quasars from the H07 mid-IR AGN sample. The H07 AGNs were identified based on the IRAC color-color
selection criterion of \cite{ster05}. For the redshift range of the H07
sample ($0.7<z\lesssim3.0$), the IRAC filters probe wavelengths at
which the characteristic power-law continuum from the reprocessed AGN
radiation starts to dominate. At these wavelengths, the light from old
stellar populations is characterized by a Rayleigh-Jeans tail of a
blackbody radiation with temperature higher than 2500K, which peaks at wavelengths different
than both the AGN accretion disk and the reprocessed AGN emission. In
addition, dust emission at near-sublimation temperature heated by the AGN is significantly stronger than that heated by
massive stars, thus AGNs can be easily identified with their mid-IR color \citep[e.g.][]{lacy04,ster05,donl12}.
These reprocessed photons at mid-IR wavelengths are less affected by obscuration
than the optical and the X-ray photons due to their much smaller
absorption cross section. Therefore, even heavily obscured AGNs
can be identified using mid-IR selection criteria.

An important feature of the H07 catalog is that the AGNs were selected
from the IRAC Shallow Survey \citep[ISS,][]{eise04}, which has
shallower flux limits than the later SDWFS survey. 
While the \cite{ster05} AGN selection 
can be contaminated by SF galaxies in deep mid-IR
surveys \citep[e.g.][]{donl12}, the contamination is negligible at the
shallow flux limits of ISS \citep[see][]{hick07obsagn,asse10sed,asse11qsolfunc}. \par

H07 also showed that the distribution of the optical ({\it R}) band to mid-IR ($4.5\micron$)
colors is bimodal for the luminous mid-IR quasars. This bimodality can be easily explained by the
difference between the SEDs of unobscured and obscured quasars. For
mid-IR selected quasars, the AGN dominates the mid-IR
wavelengths, while for obscured quasars, the nuclear emission at
optical wavelengths is heavily absorbed. Thus the SED in the {\it R} band for
obscured quasars is similar to those of normal galaxies
\citep[e.g.][and Hickox et al. 2014, in preparation]{poll06,hick07obsagn}. 
H07 has shown that the mid-IR selected quasars can be separated into two distinct populations of
quasars with an empirical color cut at ${\it R}-[4.5]=6.1$. Most ($\sim 80\%$) of the quasars in H07 with ${\it R}-[4.5]<6.1$ are spectroscopically confirmed as type 1 quasars,
therefore in this work we refer to these unobscured quasars with ${\it R}-[4.5]<6.1$  as QSO1s. 
As for quasars with ${\it R}-[4.5]>6.1$, due to the limited depth of AGES, only $\sim 5\%$ have spectroscopic measurements and can
be classified as type 2 quasars spectroscopically. However, H07 have extensively studied
the quasars with ${\it R}-[4.5]>6.1$ and have shown that these quasars are consistent with bright, obscured X-ray AGNs with $N_{\rm H}>10^{22}$
cm$^{-2}$ and with AGN bolometric luminosities $L_{\rm AGN}>10^{45}$
\ergs\ . Therefore, we refer to the obscured quasars with ${\it
  R}-[4.5]>6.1$ as QSO2s.\par

This empirical classification based on the
quasar mid-IR to optical color has been shown
to be broadly consistent with both the spectroscopically classified sources
\citep{dono13clust} and 
X-ray hardness ratio (defined as
$HR=(H-S)/(H+S)$ where H and S are the photon counts in the $2-7$ keV
band and 0.5--2 keV band, respectively) classification criteria
\citep{hick07obsagn,usma14}.
Since we lack spectroscopic confirmation for the majority of the
QSO2s, it is possible that the QSO2s are different than the optical
type 2 quasars. Recently, \cite{lacy13} have also studied the optical and near-IR spectroscopy of mid-IR
  AGNs selected using the \cite{donl12} IRAC color criteria, and found
  that a significant fraction ($\sim 33\%$) of the AGNs with evident mid-IR
  power-law continuum have no significant emission lines consistent with the optical emission line diagnostics
\citep[e.g.][]{bpt}, suggesting that these mid-IR obscured AGNs are more
deeply obscured than the optically selected type 2 AGNs. This has
also been suggested by the recent study of \cite{hain14}.\par

\subsection{The final sample}
To minimize the uncertainty in the measurements of quasar and host
galaxy properties due to the scattering in the {\it photo-z}s, 
we focus on the H11 quasar sample, which is a subsample from H07. 
The H11 quasar sample focuses on the redshift range $0.7<z<1.8$ and
limits the QSO1s to those with broad optical emission lines and robust 
{\it spec-z} measurements from AGES. For the QSOs in this redshift range, all of the
sources have ${\it I} < 18$, and the spectroscopic QSO1 sample is
highly complete. For the QSO2s in this redshift range, only 32
($\sim 9\%$) of the QSO2s have {\it spec-z} measurements. For the rest of QSO2s, we adopt
the {\it photo-z}s from \cite{brod06photoz}. This subset of QSOs
have been used in the study of clustering properties of QSO1s and QSO2s in H11, which
showed that in the redshift range $0.7<z<1.8$, the sample is highly complete at the shallow flux limits
of ISS. The mid-IR selection of this particular sample is shown to
suffer less than $<20\%$ contamination from
star-forming galaxies \citep{hick11clust}. 
Since the purpose of this work is to compare the star
  formation properties of QSO1s and QSO2s, it is important to make
  sure that any difference in SF properties is not driven by the
  presence of starburst galaxy interlopers. Also, even for bright
  mid-IR  quasars the starburst contribution in the MIR may be
non-negligible. Therefore, we rely on SED decompositions to disentangle
the AGN and starburst components. The details of the SED decomposition
are laid out at \S3. Our SED fitting results show that all of the H11
QSOs have a non-negligible ($>20\%$) AGN component. Based on the SED
fitting results, $\sim3\%$ of the H11 QSOs would have $\lagn\ $ less
than the QSO criterion ($\lagn\ >10^{45}$ erg s$^{-1}$). We therefore
limit our focus to the sample of 546 QSO1s and 345 QSO2s which
satisfies the QSO luminosity criterion and comprise
$\sim 97\% $ of the H11 sample. \par

\begin{figure}
\plotone{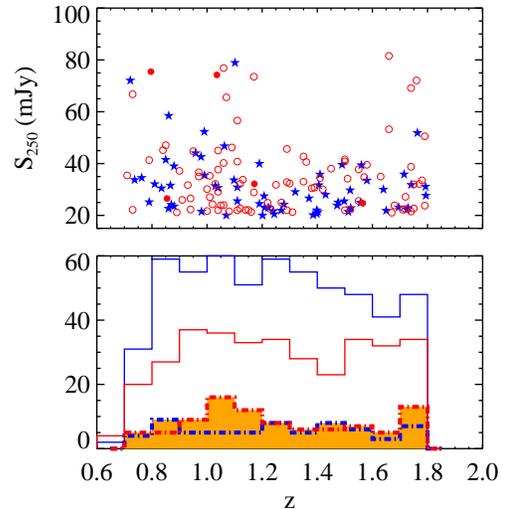}
\caption{ {\it Top}: The $250\;\micron$ distributions for QSO1s and
  QSO2s with direct SPIRE detections. QSO1s are shown as the blue
  stars. The 5 far-IR detected QSO2s with {\it spec-z} measurements are shown as the
  filled circles. The rest of the far-IR detected QSO2s with {\it photo-zs} are shown
  as open circles. {\it Bottom}: The redshift distributions of QSO1s and
  QSO2s are shown in blue and red solid lines. Also shown are the
  redshift distributions of the QSO1s and QSO2s detected at
  $250\;\micron$. The histograms for far-IR detected QSO1s and QSO2s
  are shown as the blue and red dash-dotted lines and
  filled in orange. This plot shows that the redshift distributions of
QSO1s and QSO2s are similar and QSO2s have more $250\;\micron$
detections than QSO1s.}
\label{fig:zflux}
\end{figure}

As a demonstration of the sample used in this work, we show the distribution of
  the IRAC $[3.6]-[4.5]$ to $[5.8]-[8.0]$ colors and the ${\it R}-[4.5]$ to $L_{\rm AGN}$ distributions of the quasar sample in
  Fig.~\ref{fig:sample}. We note that in Fig.~\ref{fig:sample}a, a
  small number of the H07 sources have $[3.6]-4.5$ and $[5.8]-[8.0]$
  colors that lie outside the \cite{ster05} AGN selection wedge. 
The \cite{ster05} AGN selection wedge and the H07 sample were defined using the
  original ISS catalog in which the mid-IR photometry was derived
  using the standard aperture correction, which is different than the
  PSF profile fitting corrections for individual sources in the
  \cite{brow07red} catalog. Moreover, the updated IRAC photometry
  comes from SDWFS, which is 2 times deeper than the ISS. Thus, it is not surprising that a small
  fraction ($\sim 4\%$) of the original H07 sample does not meet the
  original \cite{ster05} criterion. However, the QSO sample in
  Fig.~\ref{fig:sample}a shows a tight $[5.8]-[8.0]$ to $[3.6]-[4.5]$
  distribution, which is not seen in the similar Fig. 1a of H11 which
  use the original ISS photometry. This suggests that the majority of
  our sample does have power-law like mid-IR SED with the
  high-precision photometry of SDWFS (e.g. see Fig. 22 in H07). For the purpose
  of completeness we do not exclude the sources outside of the updated AGN selection wedge, since these
  sources might still have a non-negligible AGN component which can
  be identified by the SED fits. However, we note the
  exclusion of this small fraction of sources has little effect on the
  average properties of the QSOs. We also show the redshift and
    flux distributions of the sample in Fig.~\ref{fig:zflux}, which
    shows that the redshift distributions of the QSO1s and QSO2s are similar.

\section{SED decomposition}
Although it is now well-established that far-IR fluxes at wavelengths
longer than $60\;\micron$ in AGN host galaxies are mostly dominated by
the emission related to SF \citep[e.g.][]{netz07qsosf,mull11agnsed,rosa12agnsf}, caution is still required when estimating the SF
properties of powerful quasars. Here we use SED decompositions to
ensure that we have reliable estimates of the intrinsic AGN luminosity
($L_{\rm AGN}$) and star formation luminosity ($L_{\rm
  SF}^{\rm IR}$\footnote{Defined as the integrated $8-1000\;\micron$
luminosity of the host galaxy component only.}.) \par

To properly disentangle the emission from stars and AGN accretion, 
we fit all of the available broad band photometry in our datasets with SED
templates. 
We combine several empirically derived templates to
create AGN and host galaxy templates covering the wavelength range from optical to far-IR in our
data (see \S3.1).\par

\subsection{SED templates}
Our SED fitting procedures utilize the four
empirical SED templates spanning $0.03-30\;\micron$ described in \citet[][A10
hereafter]{asse10sed}. 
A10 have shown that the non-negative
combinations of three different galaxy templates and a single AGN
template, with the addition of extinction to the AGN component only, can
robustly describe the optical to mid-IR SEDs of a wide variety of AGNs selected using
{\it Spitzer} IRAC color \citep{asse10sed} or {\it WISE} color
\citep{asse13wise,chun14}. However, these templates do not include the far-IR
wavelengths that are crucial for this work.

To extend the A10 AGN template to the far-IR, we create {\it ad hoc}
AGN templates by replacing the hot-dust component of the A10 AGN template 
with the average infrared quasar template from \cite{netz07qsosf} and
the three infrared AGN templates (for low-, mean-, and high-luminosity AGNs)
from \cite{mull11agnsed}, which cover a wide range of AGNs with
different mid-IR to far-IR properties. 
This is based on the assumption that the SED of
the hot accretion disk around the SMBH is the same for
all the radiatively efficient AGNs, and that the differences are due
to variable extinction and different spectral shapes
in the mid- and far-IR wavelengths due to different distributions of dust. 
To account for the extinction at the AGN templates, we use the \cite{drai03} extinction law
with $R_V=3.1$, which mainly attenuates the SED at $\lambda \leq
30\;\micron$ and also produces the silicate absorption
features at mid-IR wavelengths that are common among AGN. The
extinction strength is treated as a free parameter with a range of $0<A_{\rm v}<48$. \par 

For the host galaxy templates, we consider two different
components: the contribution from the stellar population of the galaxy, which
accounts for the optical to near-IR emission; and a starburst
component, which represents the mid- to far-IR dust emission from
re-processed stellar light. For the stellar population component, we follow the approach described
in A10 by assuming that the stellar population in any galaxy is
comprised of the non-negative combination of the three empirical galaxy
templates with populations of starburst (Im),
continuous star-forming (Sbc) and old stars (elliptical), respectively \citep{asse08sed,asse10sed}. 
Unlike the elliptical template, the Sbc
and Im templates both contain hot dust components in addition to the
stellar population SEDs. Since the dust component does not extend to
the far-IR wavelengths we require and will be taken into account in
the starburst templates which we will choose later on, 
we replace the dust emission ($\lambda>4.9\;\micron$)
in the Sbc and Im templates with the SED identical to the elliptical
galaxy to create empirical stellar population templates with no dust emission. At these
wavelengths the stellar SEDs are dominated by low mass stars
similar to the elliptical template. For our mid-IR selected quasar
sample, the star light contribution is negligible at the wavelengths
where hot dust emission dominates.\par

For the starburst component, we use the 105 starburst templates from \cite{ce01} and the 64 starburst
templates from \cite{dh02}. The \cite{ce01} and \cite{dh02} starburst
templates cover a wide range of SEDs for various prototypical local star
forming galaxies, with $L_{\rm IR}$ in the range between
$10^{8}-10^{13.5} L_{\odot}$ . For SF galaxies at $z>1$, these templates have been shown to be
reliable when used to estimate SF-related $L_{\rm IR}$ with
monochromatic SPIRE observations \citep[e.g.][]{elba11ms}. 
However, as pointed out by \cite{kirk12}, the starburst galaxies at higher
redshift have different dust temperatures and SED shapes. 
Although the effect of the different dust temperature on the
estimation of $L_{\rm IR}^{\rm SF}$ is small when the wavelengths of the SPIRE bands probe the peak of the cold dust emission, it is still important to include high redshift starburst
templates to accommodate the possibly different SED shapes of high
redshift starbursts. 
Therefore, we also include the $z\sim1$ and $z\sim2$
average starburst SED templates from \cite{kirk12} in our SED fitting
analysis. We thus adopt a total of 171 starburst templates in our SED
fitting analysis. \par

Given these SED templates described above, we fit the observed photometry with an iterative $\chi^2$ minimization algorithm
(Levenberg-Marquardt) to minimize the function.

\begin{equation}
\begin{aligned} [c]
&\chi^2=\sum\limits_{i=0}^{n_{\rm filters}} \\
&\frac{F_{\rm obs,i}-a F_{\rm star,i}-b\;{\rm ext}(E(B-V))
  F_{\rm AGN,i}-c F_{\rm starburst,i}}{\sigma_i^2}.
\end{aligned}
\end{equation}
Here $F_{\rm obs,i}$ is the observed flux in the {\it i}th band and
$\sigma_i$ is its uncertainty. $F_{\rm galaxy,i},F_{\rm AGN,i}$
and $F_{\rm starburst,i}$ are the fluxes of the stellar (optical), AGN (optical to far-IR) and
starburst (far-IR) templates for filter {\it i}. The function {\rm ext} gives the extinction of the AGN flux in band {\it i} given the color excess
{\it E(B-V)}. We optimize $a, b,
c$ and {\it E(B-V)} to minimize the $\chi^2$. 

\begin{figure*}[t]
\epsscale{1.5}
\plotone{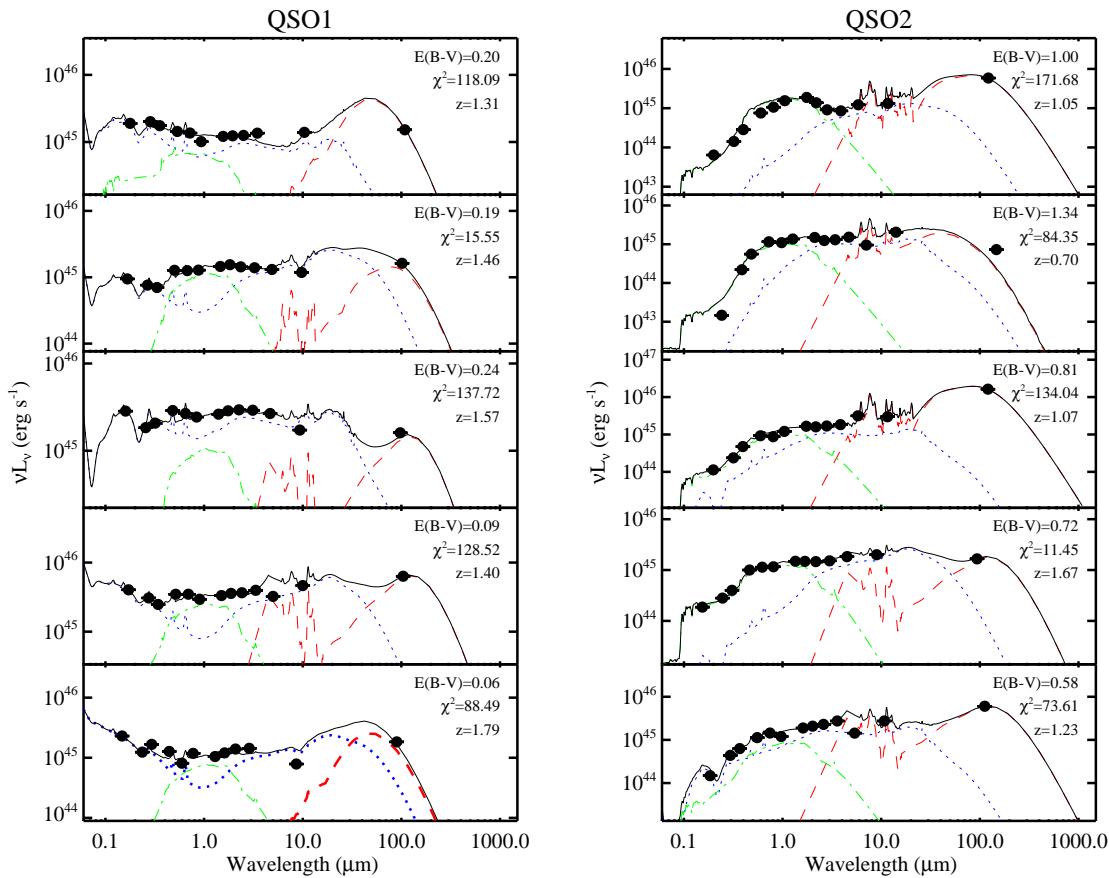}
\caption{Examples of the best-fitting SEDs (solid line) 
  in the rest-frame of each source. These sources are fitted using an
  AGN component with a \cite{drai03} extinction law (dotted curve), an empirical
  stellar component (dash-dotted) and empirical starburst templates
  (dashed). We find that almost all sources have
  an SED dominated by the starburst component in the far-IR, see \S3 for details.  }
\label{fig:SED}
\vspace{1.5ex}
\end{figure*}

\subsection{Results}
While uniform NUV to mid-IR photometric coverage is available for
all of the sources in our sample, only $\sim 18\%$ of them have a SPIRE
$250\;\micron$ detection with a flux larger than the 20 mJy. 
Since our sample is selected to have AGN-dominated mid-IR SED, the
far-IR photometry is essential to constrain the average SF
properties. For the sources with direct far-IR detections, we fit
their multiwavelength photometry with the SED templates described in
the previous section. For the 731 sources without direct far-IR
detections, we use a stacking analysis to constrain their average far-IR flux.\par

We stack the QSOs without direct far-IR detections in bins of
${\it R}-[4.5]$ (see \S4.2)  and bins of $L_{\rm AGN}$ (derived in H11
from interpolations between the IRAC photometry and the MIPS
$24\;\micron$ photometry). The uncertainty of the stacked flux is
determined by bootstrap resampling. We created 10,000 random samples by drawing
objects from the original samples with replacement until the number
of objects in each random sample is the same as the number in the original
sample. The uncertainty of the stacked flux is the
variation in the stacked fluxes of the random samples.
The details of the far-IR stacking analysis can be found in \S3.1.1 of
\cite{albe13stack}. The results of the stacking analysis are given in
Table 1. \par

To estimate the average SF luminosity of the far-IR
  non-detected QSOs, we assume that their $250\;\micron$ fluxes and
  $250\;\micron$ flux uncertainties are equal to the averages found from the stacking analysis for the sources with similar values of both
${\it R}-[4.5]$ and $L_{\rm AGN}$. We find that for the majority of the SPIRE-non detected sample, the best-fitting
SED reproduces the observed average far-IR flux well. \par

From SED fitting results with far-IR fluxes stacked in bins if ${\it
  R}-[4.5]$ and $L_{\rm AGN}$, we find that $\sim85\%$ of the quasars have a
prominent AGN hot dust component (AGN component $> 50\% $) at mid-IR wavelengths covered by the IRAC
bands. All of the quasars have a non-negligible ($>20\% $) AGN
component, confirming that the H11 sample consists of powerful
quasars. However, the AGN dominates the
quasar SED even at $24\;\micron$ for more than a half of
our sample. So the SF properties of these powerful quasars is
primarily constrained by their far-IR photometry. 
For individual sources with only the stacked SPIRE flux to anchor the starburst
component, this approach of using stacked far-IR flux can lead to
inaccurate $L_{\rm IR}^{\rm SF}$. However, the main purpose of this work is not to determine the
$L_{\rm IR}^{\rm SF}$ for individual QSOs. Instead, we take this approach to constrain the AGN
contamination in the average far-IR SF luminosity with the well-fitted mid-IR
AGN SED, which can robustly determine the average $L_{\rm IR}^{\rm
  SF}$ for different QSO populations. We note that when we compare the
stacked fluxes from either the ${\it R}-[4.5]$ bin or the $L_{\rm AGN}$ bin with the best-fitting
AGN component, we find that the average AGN contribution at
$250\;\micron$ is only $5\%$; and none of the QSOs in our sample have an AGN
component that contributes more than $40\%$ at
$250\;\micron$. Since the average intrinsic SEDs for AGNs are
expected to fall more rapidly in more luminous systems
\citep{mull11agnsed}, a starburst component is essential to reproduce
to average flux at $250\;\micron$.\par 

With the best-fitting SEDs, we calculate the total infrared luminosity for each QSO
($\ltot\ $) by integrating the best-fitting SEDs from
$8-1000\;\micron$. For each source in our sample, we also calculate
the $L_{\rm IR}$ of the starburst component by integrating the host galaxy
component of the best-fitting SED over the same wavelength range
($L_{\rm IR}^{\rm SF} $ hereafter) \footnote{For clarification, we use $L_{\rm
    IR}$ as a general terminology for any integrated $8-1000\;\micron$
  luminosity. $\ltot\ $ is the integrated $L_{\rm IR}$ of the 
  best-fitting SED and $L_{\rm IR}^{\rm SF}$ is the $L_{\rm IR}$ of the host galaxy component only.}.
 We find that for our sample, the average ratio between $L_{\rm IR}^{\rm SF}$ and $\ltot\ $ is $\sim 55\%$, indicative of a non-negligible
 AGN contribution to $\ltot\ $ in these mid-IR luminous quasars. This 
 is also consistent with the \cite{kirk12} results of the study combining {\it
   Spitzer} IRS spectra and multi-band {\it Herschel} photometry for
 $z\sim 1$ and $z\sim 2$ ULIRGs.
However, the far-IR flux probed by the SPIRE
$250\;\micron$ band is dominated ($>70\%$)  by the starburst
component for $\sim 90\%$ of our sample. This result confirms that quasar SEDs are still
dominated by a starburst component at rest-frame wavelengths longer than
$100\;\micron$ \citep[e.g.][]{netz07qsosf,mull11agnsed,rosa12agnsf,delm13}. 
Therefore, the average $L_{\rm IR}^{\rm SF}$ of powerful mid-IR
quasars can still be robustly measured with the inclusion of far-IR
photometry.\par

\begin{figure}
\epsscale{1.1}
\plotone{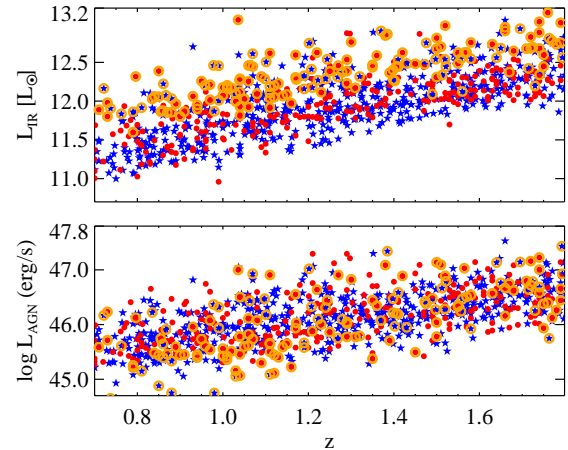}
\caption{The redshift distributions of the ($8-1000\;\micron$) $L_{\rm
    IR}^{\rm total}$ and the
  AGN bolometric luminosity derived from SED fitting and far-IR
  stacking for non-detected sources described in \S3.2.
The symbols and lines represent the same
  subsets of objects shown in Fig. 1, where blue stars and red
  circles are QSO1s and QSO2s, respectively. Individually
  far-IR detected sources (FIR-QSO) are enclosed with orange
  circles. }
\label{fig:qso-zdist}
\end{figure}

We next derive the bolometric AGN luminosity ($L_{\rm AGN}$) by directly integrating the de-absorbed
AGN component. We compare the $L_{\rm AGN}$ from our SED fitting with
the $L_{\rm AGN}$ derived in H11, and found that our $L_{\rm AGN}$ is
higher than the H11 $L_{\rm AGN}$ by an average of  $\sim 0.1$ dex. 
Since the derivation of $L_{\rm AGN}$ in H11 did not correct for the weak
but non-negligible dust obscuration and SF galaxy contamination in the mid-IR, it is not surprising that our
$L_{\rm AGN}$ is different. However, the difference is small due to the fact
that most of the quasars in this sample are dominated by the AGN and
the dust obscuration only weakly attenuates the SED at mid-IR
wavelengths. We show the redshift distributions of $\ltot\ $ and $L_{\rm AGN}$ in
Fig.~\ref{fig:qso-zdist}. \par

\section{Far-IR observations and quasar obscuration}
Several studies have pointed out that even for AGNs with quasar-like luminosites, their far-IR
SEDs are often dominated by the cold dust emission heated
by young stars \citep[e.g.][]{netz07qsosf,lacy07,kirk12,mull12agnsf}. 
Although alternative cases have been reported 
in some recent studies \citep[e.g.][]{hine09,dai12}, the cold temperature
of the far-IR emission still requires the AGN-heated dust to reside at
a large distance from the central
SMBH \citep{sand88}. In addition, powerful molecular outflows
  might also produce strong far-IR emission
  \citep[e.g.][]{sun14outflow}. However, the number of 
  AGNs and starburst galaxies with confirmed warm molecular outflows
  from high-resolution far-IR or sub-mm observations is
  limited. Nonetheless, strong far-IR emission for quasars implies the
  existence of dust extending to a large distance well beyond the
  putative obscuring torus, and a comparison of far-IR properties
  between QSO1s and QSO2s could determine that whether the quasar obscuration is related to the large-scale dust.

\subsection{SPIRE detection fraction}
We begin with a very simple test 
by measuring the far-IR detection fraction above a flux limit of $20$ mJy
in the {\it Herschel} SPIRE $250\;\micron$ filter ($f_{250}$ hereafter) for QSO1s and
QSO2s separately. In our survey region, the shallowness
of the SPIRE observation implies that any quasars in our sample with
a far-IR detection are at least as bright as luminous infrared
galaxies (LIRGs, which are defined as galaxies with $L_{\rm IR}(8-1000\;\micron) > 10^{11}L_\odot$ ) at the same
redshifts. At first glance, we find that among the 546
QSO1s, only 68 ($12\%$) of them are detected in the SPIRE $250\;\micron$
filter, while 102 of the 345 ($29\%$) QSO2s have far-IR
detections. \par

\begin{figure}
\epsscale{1.1} 
\plotone{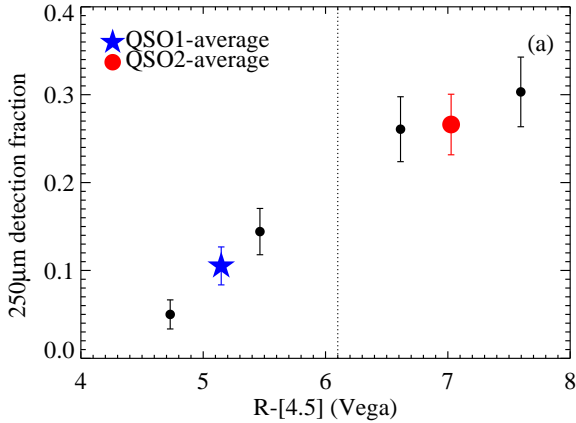}
\caption{ The $250\;\micron$ detection fraction ($f_{250}$) as a function of the ${\it R}-[4.5]$
  color. The $f_{250}$ for all QSO1s is shown as the blue star, and
  the $f_{250}$ for all QSO2s (weighted
  to have a redshift and AGN bolometric luminosity distribution similar to
  that of QSO1s) is shown as the red circles. This figure shows that
  QSO2s are 2.4 times more likely to have a far-IR detection.}
\label{fig:r45}
\end{figure}

We next use ${\it R}-[4.5]$ as a rough proxy of the obscuration strength on
the nuclear emission and study whether $f_{250}$ is related to the
extinction at the {\it R} band.
The primary goal of this analysis is to study whether the star
formation properties of the quasar-hosting galaxies are related to the
observed obscuration. Therefore, it is important to make sure the
QSO1s and QSO2s are matched in key properties, i.e. redshift
and AGN bolometric luminosity ($L_{\rm AGN}$, estimated from the
  SED fitting analysis discussed in \S3).
We divide our sample into four bins of ${\it R}-[4.5]$ and calculate 
$f_{250}$ for each bin by applying statistical weights to each quasar so that the samples are matched in
redshift and $L_{\rm AGN}$ in each bin.
We find that the weighted $f_{250}$ increases rapidly with ${\it R}-[4.5]$ for
QSO1s. For QSO2s, $f_{250}$ is only weakly correlated with ${\it
  R}-[4.5]$, but the $f_{250}$ of QSO2s is higher than that of QSO1s by a
factor of $2.7\pm0.2$. We show the $f_{250}$ to ${\it R}-[4.5]$
relation in Fig.~\ref{fig:r45}.

\subsection{The average SPIRE $250\;\micron$ fluxes}
Due to the flux limit of the SPIRE catalog, $f_{250}$
only reflects the fraction of QSOs hosted by galaxies with very luminous far-IR
emission. To extend this analysis, we estimate the average
$250\;\micron$ flux in bins of ${\it R}-[4.5]$ to test if the far-IR
flux of our quasar sample also evolves with ${\it R}-[4.5]$. 
The measure the average $250\;\micron$ flux ($S_{250}$) for all QSOs, we first
estimate the average $250\;\micron$ flux for the sources without direct SPIRE
observations using the stacking analysis discussed in \S3.2. 
Next, the stacked fluxes is combine with the sources individually
detected in far-IR to calculate the average $S_{250}$ for all QSOs. 
We show the average $250\;\micron$ flux ($S_{250}$) in the same ${\it R}-[4.5]$
bins used for Figure ~\ref{fig:r45}, with results given in Table 1.\par

From the stacking analysis,
we find that for QSOs without direct far-IR detections, 
there is no significant dependence of the average $S_{250}$ on ${\it
  R}-[4.5]$. However, the mean $S_{250}$ of the individually
detected far-IR sources shows a ${\it R}-[4.5]$
dependence similar to that of $f_{250}$. Therefore, 
driven by the increasing fraction of far-IR luminous sources, the {\it average} $S_{250}$ of
the entire QSO sample is correlated with ${\it R}-[4.5]$. 

This result shows that for powerful mid-IR quasars with more dust attenuation at
optical wavelengths, the average far-IR emission is stronger. This
implies that at least part of the obscuration seen in the obscured
QSOs might be due to the far-IR emitting dust. 

\begin{table*}
\label{table:stacking}
\centering
\caption{Results of Far-IR Stacking Analysis}
\begin{tabular}{ccccc}
\toprule
& &Average $f_{250}$ in bins of R-4.5 & &\\
\tableline
    ${\it R}-[4.5]$ (Vega)              & 4.7         & 5.4         & 6.6         & 7.5         \\
    $S_{250}$(mJy) (ND)             & $6.58\pm0.73$ & $7.93\pm0.59$ & $7.39\pm0.75$ & $8.57\pm1.00$ \\
    $S_{250}$(mJy) (All)            & $8.04\pm0.84$ & $11.4\pm0.77$ &     $16.4\pm1.50$ & $18.0\pm0.21$ \\
    $\langle z\rangle$ (all)             & 1.31 & 1.20 &   1.19 & 1.31 \\
\toprule
& &Average $f_{250}$ in bins of $L_{\rm AGN}$ & &\\
\tableline
    $\log\langle L_{\rm AGN}\rangle$ [erg $s^{-1}$] & 45.2        & 45.6        & 46.1        & 46.5        \\
    $S_{250}$(QSO1)(mJy)         & $3.76\pm0.82$ & $5.49\pm0.70$ & $5.21\pm0.75$ & $7.70\pm0.95$ \\
    $S_{250}$(QSO2)(mJy)      & $3.56\pm0.94$ & $8.84\pm1.40$ &    $5.18\pm0.66$ & $8.44\pm2.70$ \\
    z (average)             & 0.98 & 1.17 &     1.36 & 1.49 \\

\tableline
\end{tabular}
\tablecomments{Results from the far-IR stacking analysis described in \S4.2. The top half of
  the table shows the results for sources binned in
  ${\it R}-[4.5]$. The first row shows the results for the far-IR
  non-detected (ND) sources only, and the second row shows the
  result for the entire QSO sample. The average
  redshift for all QSOs in each bin is also given. The bottom half of the table shows the results for the entire QSO1 and
  QSO2 samples binned in their AGN bolometric luminosity. The average
  redshift for each ${\it R}-[4.5]$ and $L_{\rm AGN}$ bin is also
  listed; we note that there is only a $<0.02$ average redshift
  difference between the QSO1s and QSO2s.}
\end{table*}

\begin{figure*}[t]
\epsscale{1.15} 
\plottwo{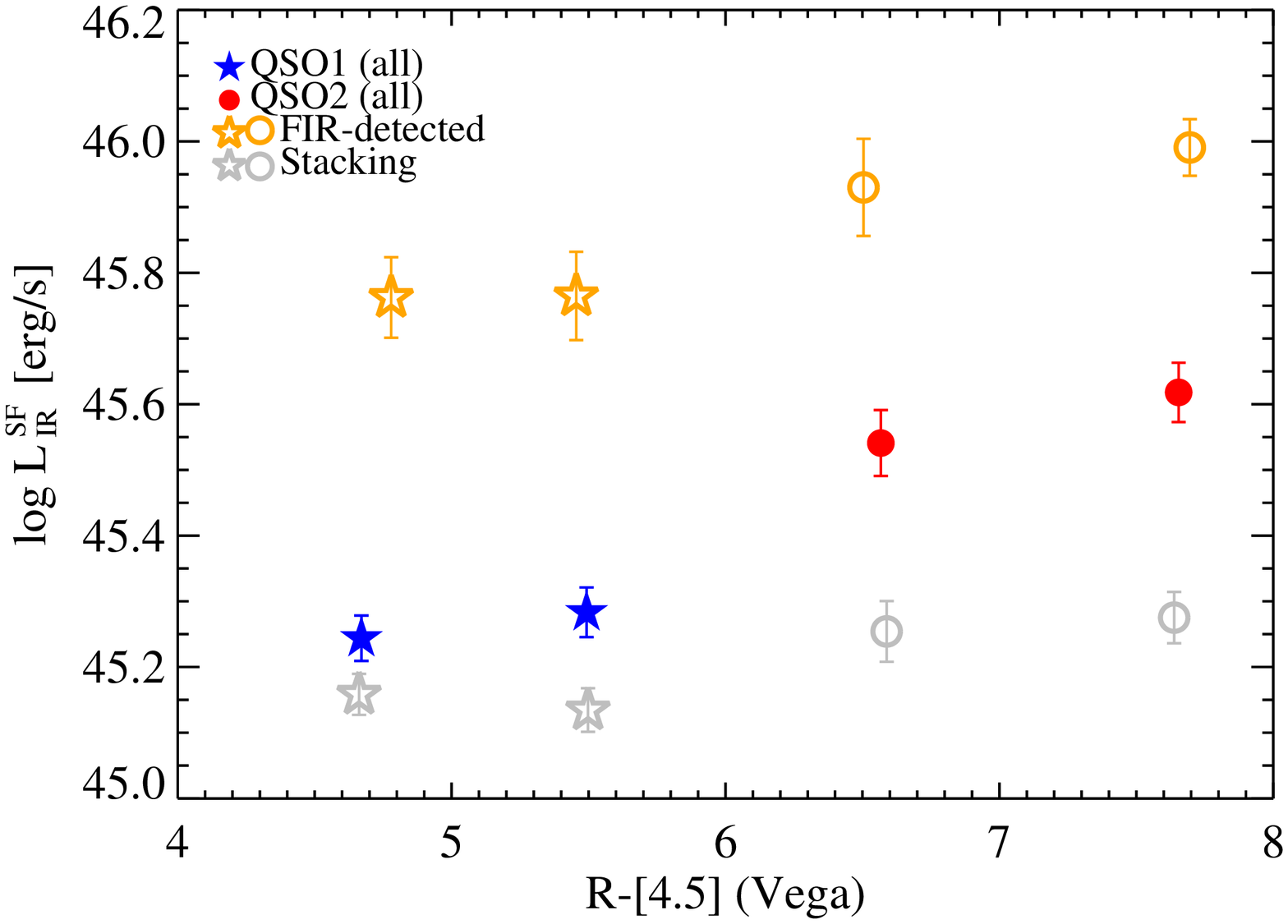}{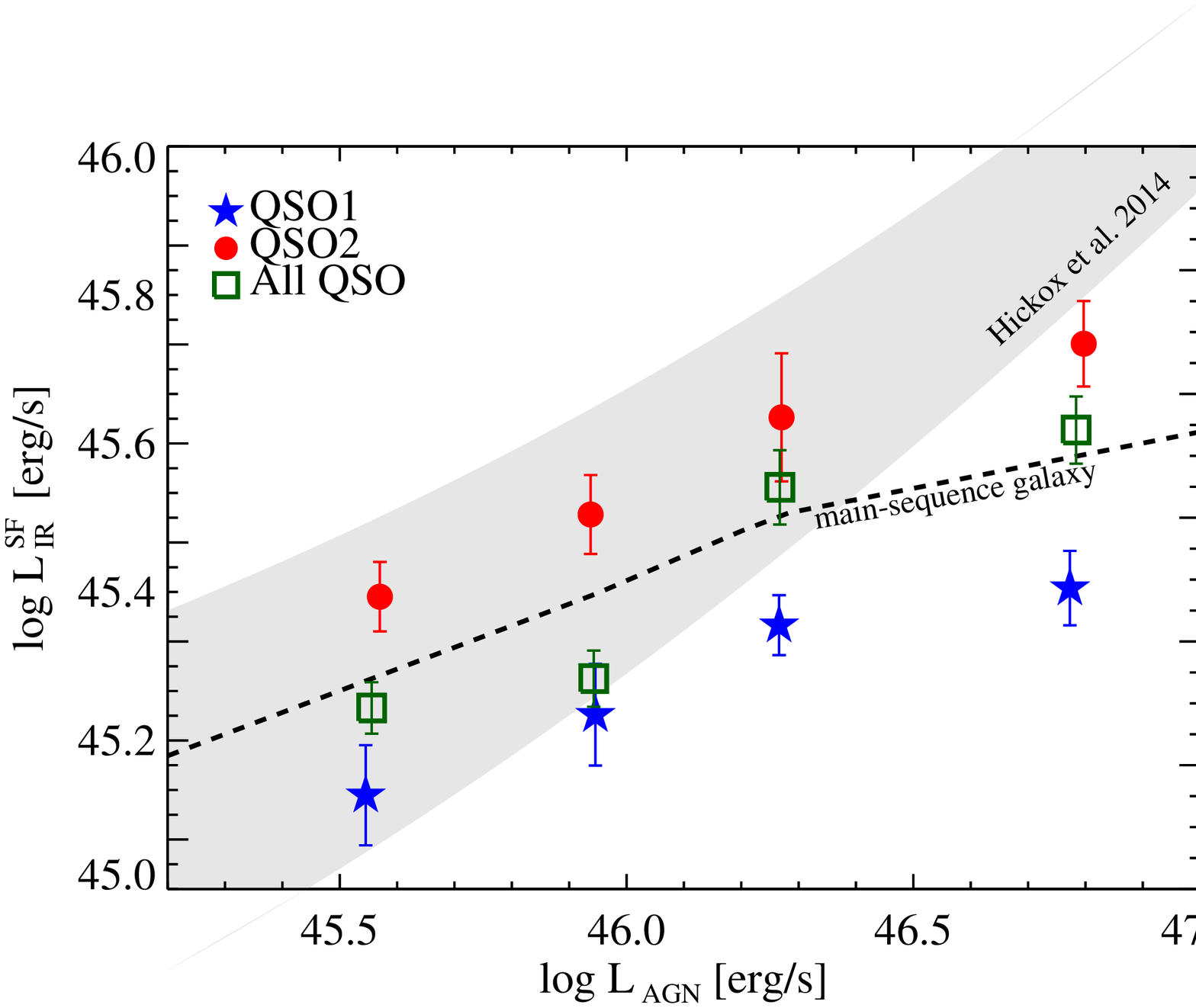}
\caption{(a) {\it Left}: The relationship between
  $L_{\rm IR}^{\rm SF}$ and ${\it R}-[4.5]$ for the mid-IR selected
  QSOs. Far-IR detected QSOs are shown by the orange, open symbols and
  the stacking results for the far-IR non-detected QSOs are shown by
  the gray, open symbols. In both cases, there is only marginal
  increase in $L_{\rm IR}^{\rm SF}$ with ${\it R}-[4.5]$. However, driven by the much
    higher far-IR detection fraction in QSO2s, the average $L_{\rm IR}^{\rm SF}$ for QSO2s is
  higher than the average $L_{\rm IR}^{\rm SF}$ for QSO1.
(b) {\it Right}: the average $L_{\rm IR}^{\rm SF}$ for QSOs in bins of $L_{\rm AGN}$. The
  results for QSO1s are shown as the filled, blue stars; while the results
  for QSO2s are shown as the filled red circles. For the entire QSO
  population, we also show their average $L_{\rm IR}^{\rm SF}$ in four bins of
  $L_{\rm AGN}$ as the open, green squares. For comparison, we
  show the \cite{hick14agnvar} model evaluated at $0.7<z<1.8$ as the
  shaded region. We note that the average redshift in each bin
  increases with $L_{\rm AGN}$; and the relationship between $\langle
  L_{\rm IR}^{\rm SF}\rangle$ and $L_{\rm AGN}$ is consistent with
  the redshift evolution for main-sequence SF galaxies with
  $M_\star=10^{11}M_\odot$. We show the redshift evolution of MS galaxies as the dashed line. The $L_{\rm IR}^{\rm SF}$ for
all QSOs coincides with the evolution of MS galaxies, suggesting a
connection between QSO host galaxies and MS star-forming galaxies (see \S5.2).}
\label{fig:lagn}
\vspace{1.5ex}
\end{figure*}

\section{The average $L_{\rm IR}^{\rm SF}$ of QSO1s and QSO2s}
In \S 3, we found that QSO2s have higher $f_{250}$ and $S_{250}$
than QSO1s. In this section, we examine whether the observed difference in the average far-IR fluxes
of QSO1s and QSO2s can be associated with the difference in the star
formation properties of their host galaxies, as predicted by the galaxy evolution
models with a connection between quasar obscuration and star formation. \par

We first calculate the median star formation luminosity by averaging
the $L_{\rm IR}^{\rm SF}$ derived from SED fitting (\S4). We find that
for QSO1s, the mean $L_{\rm IR}^{\rm SF}$ is $10^{45.29 \pm0.03}$ \ergs\
($10^{11.71} L_\odot$); and for QSO2s, the mean $L_{\rm IR}^{\rm SF}$
is $10^{45.59\pm0.04}$ \ergs\ ($10^{12.01} L_\odot$).
We find that similar to $f_{250}$ and the average $S_{250}$,
  the median $L_{\rm IR}^{\rm SF}$ for QSO2s is significantly higher than
  that of QSO1s by $0.30$ dex. The higher $L_{\rm IR}^{\rm SF}$ in QSO2s confirms the results from \S3 that QSO2s are
associated with host galaxies with more star-forming cold dust which might be
obscuring the nuclear emission. \par

\subsection{$L_{\rm IR}^{\rm SF}$ vs. ${\it R}-[4.5]$}
To further study the relation between AGN obscuration and $L_{\rm
  IR}^{\rm SF} $, we divide our QSO sample into bins of ${\it
  R}-[4.5]$ as a proxy for AGN obscuration. To be consistent with \S3.1 and avoid the
uncertainties by the lack of {\it spec-z}s of QSO2s, we opt to use ${\it R}-[4.5]$ instead of the
$E(B-V)$ derived from SED fitting as the proxy for nuclear
obscuration, since the uncertainties of the \pz\ limit the accuracy of
our ability to measure $E(B-V)$. 
We calculate the median AGN contribution at observed
frame ${\it R}$ and $[4.5]$ bands using the best-fitting SEDs in each bin. 
We find that towards redder ${\it R}-[4.5]$ colors, the AGN contributes
$89\%, 79\%, 29\%, 12\%$ at the {\it R} band and $74\%, 59\%, 43\%, 60\%$
at the $[4.5]$ band. This suggests that the empirical ${\it R}-[4.5]$ color
is indeed a good proxy of AGN obscuration at optical wavelengths,
and that the QSO classification based on the ${\it R}-[4.5]$ color is
reliable. We then calculate the average $L_{\rm IR}^{\rm SF} $ ($\langle L_{\rm IR}^{\rm SF}\rangle$) in bins
of ${\it R}-[4.5]$. We again estimate the uncertainty
in $\langle L_{\rm IR}^{\rm SF}\rangle$ using bootstrap
resampling. 

We show the results in Fig.~\ref{fig:lagn}a. Even though
the $L_{\rm IR}^{\rm SF}$ difference between QSO1s and QSO2s is large,  $\langle L_{\rm IR}^{\rm SF}\rangle$ does not
show the strong dependence on ${\it R}-[4.5]$ seen for $f_{250}$ and the
average $S_{250}$ (Fig.~\ref{fig:lagn}a.) within QSO1s and QSO2s. 
Also, the $L_{\rm IR}^{\rm SF}$ difference between the far-IR detected QSO1s and
QSO2s and the far-IR non-detected QSO1s and QSO2s are only $0.14$ dex
and $0.17$ dex, which are not as significant as the $0.30$ dex
difference between all QSO1s and QSO2s. 
Thus the higher $\langle L_{\rm IR}^{\rm SF}\rangle$ of the QSO2 sample is mainly driven by its higher fraction of far-IR
luminous SF galaxies. \par

\subsection{$L_{\rm IR}^{\rm SF}$ vs. $L_{\rm AGN}$}
To explore the connections between AGN and host galaxy growth rates in
different QSO populations, we separately divide the QSO1s and QSO2s into bins of $L_{\rm
  AGN}$, and measure the $\langle L_{\rm IR}^{\rm SF}\rangle$ for each
bin. The uncertainties are again estimated by
bootstrapping as discussed in \S5.1.
We plot the $L_{\rm AGN}$-$L_{\rm IR}^{\rm SF} $ relations for QSO1s and QSO2s in
Fig.~\ref{fig:lagn}b. We find that both the QSO1 sample and the QSO2 sample have a positive
$L_{\rm IR}^{\rm SF}$-$L_{\rm AGN}$ correlation with similarly shallow slopes,
with $\log L_{\rm IR}^{\rm SF} \propto 0.25 \log L_{\rm AGN}$ for
QSO1s and $\log L_{\rm IR}^{\rm SF} \propto 0.27 \log L_{\rm AGN}$ for
QSO2s.  However, the $\langle L_{\rm IR}^{\rm SF}\rangle$ for QSO2s is
higher than that of QSO1s in each $L_{\rm AGN}$ bin by 0.28,0.25,0.32 and
0.28 dex. 

For comparison, we also show the $L_{\rm AGN}-L_{\rm IR}^{\rm SF}$ model from
\cite{hick14agnvar} which assumes a direct, linear correlation
between the {\it average} $L_{\rm AGN}$ and $L_{\rm IR}^{\rm SF}$ while the observed $L_{\rm AGN}$ is modulated by
rapid variability. We calculate the \cite{hick14agnvar} relation spanning the
$0.7<z<1.8$ redshift range in Fig.~\ref{fig:lagn} as the shaded region.
We find that for all QSO2s besides those in the most luminous $L_{\rm AGN}
$ bin, our result is consistent with the H14 model, and the QSO1s have $L_{\rm IR}^{\rm SF}$ slightly lower than the model
predictions throughout the entire $L_{\rm AGN}$ range. However, for both types of QSOs, the slope of the $L_{\rm
  IR}^{\rm SF}$-$L_{\rm AGN}$ relation appears to be shallower than
the prediction of the simple model if we take the different average
redshift in each $L_{\rm AGN}$ bins into account.\par

As was shown by the results in Table 1, the average redshifts for more luminous QSOs
are higher. Although our far-IR stacking approach can reliably measure
the average far-IR fluxes and hence the average $L_{\rm IR}^{\rm SF}$, it is
important to note that the difference in the average redshift between the
lowest and the highest $L_{\rm AGN}$ bins is 0.5, and it is possible that
the observed correlation between $L_{\rm IR}^{\rm SF}$ and $L_{\rm AGN}$ is partially
driven by the cosmic evolution of star formation and AGN accretion. Even though recent studies of the evolution of cosmic infrared luminosity density have
shown that for the redshift range of the four $L_{\rm AGN}$ bins in
Fig.~\ref{fig:lagn}b, the cosmic infrared luminosity is only weakly
correlated with redshift \citep[$\rho_{\rm IR}\propto
(1+z)^{-0.3\pm0.1}$ for $1.1<z<2.85$,][]{grup13}, there is still a
substantial evolution in the specific SFR (SFR per unit stellar mass) of ``main sequence (MS)'' SF galaxies
\citep[e.g.][]{elba11ms}.\par 

A simple estimate of the effect of MS
galaxy SFR redshift evolution can be made by calculating the average
$L_{\rm IR}^{\rm SF}$ for MS galaxies with $M_\star=10^{11}M_\odot$ at the redshift
range of our QSO sample. From the theoretical framework of
  dark matter halo abundance matching methods \citep[e.g.][]{behr13}, $10^{11}M_\odot$ is the typical galaxy stellar mass
hosted by dark matter haloes of mass $M_{halo}=10^{13.1}[h^{-1}M_{\odot}]$, which is the halo mass reported
  in recent clustering studies of mid-IR QSOs
  \citep[e.g.][]{hick11clust,dono13clust,dipo14}. At the redshift of
  each QSO in our sample, the $L_{\rm IR}^{\rm SF}$ for a $10^{11}M_\odot$ MS galaxy can therefore be evaluated using the
  redshift-dependent SFR-$M_\star$ relation from \cite{whit12} and a
  Kennicutt relation, $SFR=1.09\times L_{\rm IR}^{\rm SF}/L_\odot$
  \citep[][modified for a Chabrier initial mass function]{kenn98araa}. We find that the average $L_{\rm IR}^{\rm SF}$ for MS
  galaxies evaluated using this approach is consistent with the
  average $L_{\rm IR}^{\rm SF}$ in each $L_{\rm AGN}$ bin. This implies that for mid-IR quasars in this redshift range, the
$L_{\rm IR}^{\rm SF}$ and instantaneous $L_{\rm AGN}$ might not be directly connected but still follow
similar redshift evolution. This suggests that a common physical parameter
(e.g. a common gas supply) might be driving the evolution of both
SMBHs and galaxies. 
This is also consistent with the recent study of
  \cite{rosa13sfagn}, which showed that the $L_{\rm IR}^{\rm SF}$-$L_{\rm AGN}$
  correlation for broad emission line QSOs is consistent with a
  scenario in which quasars are hosted by normal star-forming
  galaxies. \par

For mid-IR selected AGNs, contamination from star-forming galaxy
interlopers is also an issue that might introduce biases to the
measurements of AGN and star formation luminosities. 
While our SED fitting results show that $95\%$ of
the sources in our sample are indeed powerful QSOs with AGN-dominated
mid-IR SED, it is still very important to verify that the observed $L_{\rm IR}^{\rm SF}-L_{\rm
AGN}$ relation in Fig.~\ref{fig:lagn} is not caused by the
incompleteness of our AGN selection criterion. In particular,
a very powerful SF galaxy can harbor a heavily obscured AGN and still
have starburst-like mid-IR SED. In such
case, the IRAC color of the SF galaxy would fall {\it out} of the \cite{ster05} selection wedge
\citep[e.g.][]{kirk12,asse13wise,chun14}, and the sample selected with
the \cite{ster05} wedge would show a biased $L_{\rm IR}^{\rm SF}-L_{\rm
  AGN}$ relation. \par

By happenstance, the mid-IR color of typical star-forming galaxy
templates occupies the lower-left corner of the \cite{ster05} wedge at
the redshift range of this sample \citep[e.g.][Fig. 17]{donl12}. Therefore, any AGN contribution in
addition to the SF galaxy templates would easily promote a SF
galaxy into the AGN identification wedge. This is particularly true
for our sample of luminous quasars. We test this effect by first
normalizing the $6\;\micron$ luminosity for AGN templates we used in
the SED fitting described in \S4 to $L_{\rm AGN}=10^{45}$ \ergs\ to meet
the minimum $L_{\rm AGN}$ criterion of our QSO sample. We next combine these
normalized AGN templates with two archetypical starburst galaxy templates, M82 and
Arp 220 from \cite{poll07agnsed}. We find that to have a star forming galaxy falling out of the AGN identification wedge while having an
$L_{\rm AGN} \sim 10^{45}$ erg s$^{-1}$, its $L_{\rm IR}^{\rm SF}$ must exceed
$10^{47}$ erg s$^{-1}$. This suggests that an SF galaxy must be a
hyper-luminous infrared galaxies (HyLIRG, $L_{\rm
  IR}>10^{13}L_{\odot}$) to be able to hide an
underlying quasar-like AGN component. According to the far-IR
luminosity function of \cite{grup13}, there would only
be less than 2 HyLIRGs in the redshift range of our
sample given the size of the \bootes\ survey region. Because of the large number of sources in the
lowest $L_{\rm AGN}$ bin of our sample, the inclusion of the 2 additional QSOs hidden in HyLIRGs would only affect the average $L_{\rm IR}^{\rm SF}$ by
0.15 dex, which is still within the 2 $\sigma$ range of the original $L_{\rm
  SF}$. \par

We can also estimate biases in $L_{\rm IR}^{\rm SF}$ due to the sources not
selected in the AGN identification wedge by examining the
SEDs of the sources not classified as mid-IR QSOs. In the redshift range of our sample, there are
77 SPIRE-detected sources outside the
\cite{ster05} wedge. When we fit the SEDs of these sources we find that most of these powerful SF galaxies have no significant
AGN contribution in the mid-IR. For the sources without a mid-IR AGN component, we assume that AGN
contributes $<10\%$ to the mid-IR monochromatic luminosity at
$6\;\micron$ as an upper limit. Only 4 of the 77 have $L_{\rm AGN}$ satisfying the $L_{\rm
AGN}>10^{45}$ erg s$^{-1}$ criterion. If we add these four sources
into the lowest $L_{\rm AGN}$ bin, the average $L_{\rm IR}^{\rm SF}$ would only
increase by 0.05 dex. Therefore, we conclude that the $L_{\rm
  SF}-L_{\rm AGN}$ correlation observed in Fig.~\ref{fig:lagn} is not
biased by the exclusion of heavily obscured quasars hidden in
powerful SF galaxies.\par

\begin{table*}
\label{table:number}
\centering
\caption{Number of X-ray detected sources}
\begin{tabular}{lcccccccccc}
\toprule
          & $N_{X}$         & $N_{NX}$ & $f_X$& $\langle z \rangle$ & $\log\langle L_{6\micron}\rangle[{\rm erg s}^{-1}]$& $\langle\log L_{\rm IR}^{\rm SF}\rangle
          [L_\odot]$ & $\langle L_X \rangle$[erg s$^{-1}$] &
          $\langle A_{\it V}\rangle$\\
\hline
    QSO1 (ALL)            & 356 & 190 &  $0.65$ & 1.24& 46.20
    & 11.74 & $44.17\pm0.03$  & 0.32 \\
    QSO1 (FIR)            & 30 & 35 &  $0.46$   & 1.26& 46.38 & 12.21 & $43.93\pm0.09$ & 0.50 \\
    QSO1(No FIR)         & 326 & 155 &    $0.67$  &1.24 & 46.17 & 11.62 & $44.19\pm0.03$ & 0.28\\
\hline
    QSO2 (ALL)  & 119 & 235 &  $0.34$  &1.26& 46.25& 12.06 & $43.78 \pm0.05$ & 2.62\\
    QSO2 (FIR)            & 23 & 74 &  $0.24$ & 1.24&46.27&12.34  & $43.66\pm0.10$ & 2.85 \\
    QSO2(No FIR)         & 87 & 161 &    $0.35$& 1.26 &46.24&11.80  &$43.82\pm0.05$ & 2.59\\
\hline
\end{tabular}
\tablecomments{The number counts and properties of different subsamples of QSOs
classified based on the X-ray and far-IR detections. The
properties listed here are the number of sources with X-ray detections
($N_X$), the number of source without X-ray detections ($N_{NX}$),
the X-ray detection fraction ($f_X$), the average redshift ($\langle
z\rangle$), the average AGN bolometric luminosity ($\langle\log L_{\rm AGN}\rangle$, see \S4), the
rest-frame 2--10 keV X-ray luminosity and $A_{\it V}$ (see \S4 and \S5.3). }
\end{table*}

\section{X-ray properties of mid-IR QSOs}
In \S5, we have shown that for mid-IR selected QSOs, the AGN
obscuration at optical wavelengths and the AGN mid-IR luminosity
can both be connected to the star formation of their host galaxies.
Here we use an alternative AGN accretion rate indicator, the
X-ray emission, to study the interplay between the far-IR emitting
dust and the X-ray emission. \par

We first count the number of detections in the far-IR and the X-ray
observations. We note that the counts of X-ray detections can be affected by the varying sensitivity
of the {\it Chandra} observations across the field
\citep[e.g.][]{mend13}. However, the \xbootes\ observations are
relatively uniform in depth \citep[$\sim 4-8\times10^{-15}$ \ergs\ ,][]{kent05} and this effect should be
equivalent for both QSO types. The results of the detection fractions
are summarized in Table 2. 
First, QSO1s have an X-ray detection fraction of
$65\%$, which is much higher than the $12\%$ far-IR detection
fraction for QSO1s (see \S3). For QSO2s, the X-ray detection fraction is only $34\%$ while
the far-IR detection fraction is $27\%$. 
We also find that the presence of far-IR emission is associated with lower
X-ray detection fractions in both QSO1s and QSO2s. 
For QSO1s, $91\% $ of the X-ray detected AGNs are not detected in the
far-IR; while for QSO2s, $73\% $ of the X-ray AGNs have no detectable
far-IR emission. These number counts imply that large scale dust might also play an important role
in the absorption of X-rays. \par

Since a large fraction ($\sim 76\%$) of the far-IR detected QSO2s
have no direct X-ray detection, these far-IR bright
QSO2s might not be included in an X-ray selected AGN sample. 
However, since both the far-IR and the X-ray
observations in our sample are relatively shallow, the
detection fractions only reflect the incidence of bright starburst
and bright X-ray AGNs. 
To compensate for the shallow flux limit of the \xbootes\ survey,
we use an X-ray stacking analysis to estimate the average X-ray
luminosity for the sources without direct X-ray detections.
We defined the stacked X-ray counts as the average
number of background-subtracted photons detected within the 90\%
point-spread function (PSF) energy encircled radius at 1.5 keV,
$r_{90}$, where $r_{90}=1''+10''(\theta/10')^2$. 
Here $\theta$ is the off-axis angle from the {\it Chandra} optical
axis\footnote{{\it Chandra} Proposers's Observatory Guide (POG),
  available at \url{http://cxc.harvard.edu/proposer/POG}.}. 
We adopt 3.0 and 5.0 counts s$^{-1}$ deg$^{-2}$ for the diffuse X-ray
background at 0.5--2 keV and $2-7$ keV
bands, which closely matches the average background in annuli after
excluding detected X-ray sources. We refer to \S5.1.1 of H07 and
\cite{chen13sfagn} for details about X-ray stacking analysis.\par

We performed the analysis on the subsamples listed in
Table 2. To calculate the average X-ray luminosity for each subsample of QSOs, 
we first calculate the rest-frame 2--10 keV $L_{\rm X}$ for each X-ray
detected quasar from their observed 0.5--7 keV
luminosity with a {\it k}-correction using the spectral index derived from the
flux ratio between the soft (0.5--2 keV) and the hard ($2-7$ keV) bands. 
In detail, the X-ray spectrum was assumed to be a simple
power-law and the hardness ratio was used to determine the power-law
  index ($\Gamma$) using PIMMS and the ACIS Cycle 4 on-axis response
  function. For the X-ray detected QSO1s, the average index is
  $\Gamma=1.83$, which is typical for unabsorbed AGN X-ray
  spectra. For the X-ray detected QSO2s, the average index of $\Gamma=1.24$ is
  consistent with AGNs with moderate absorbed spectra
  \citep{hick07obsagn}. 
With the $\Gamma$ derived from the observed
  hardness ratio, we {\it k}-corrected the observed-frame 0.5--7 keV
  luminosity to the rest-frame 2--10 keV, without de-absorbing the
  X-ray spectra. For the X-ray non-detected sources, we calculate their average
rest-frame 2--10 keV $L_{\rm X}$ by {\it k}-correcting the stacked
0.5--7 keV luminosity derived from the stacked X-ray flux using the median redshift and the hardness
ratio of stacked X-ray photons. The general X-ray properties of QSO1s
and QSO2s have been discussed at length in \S 5 of \cite{hick07obsagn}.
Combining the $L_{\rm X}$ of the individually X-ray detected
sources with the average $L_{\rm X}$ for the non-detected sources from
stacking, we obtain the average X-ray luminosities in Table
2. These X-ray luminosities are the observed values and are
  not corrected for absorption.
We also list the median extinction $A_{\it V}$ of the AGN template at optical wavelengths from
the SED fitting results (see \S4). 
For heavily obscured AGNs, the SED at optical wavelengths can be
dominated by the host galaxy, and the average $A_{\it V}$ may be a lower
limit. However, we do find a much larger value of $A_{\it V}$ for far-IR detected QSO1s than that for far-IR
non-detected QSO1s, which implies that the presence of star-forming dust attenuates
both the optical and X-ray AGN emission. 
\par

We can also use the total numbers of stacked hard-band ($2-7$ keV) and
soft-band (0.5--2 keV) photons to obtain the hardness ratio for different QSO subsamples.
We find that for QSO1s, the HR remains a constant $HR\sim -0.42$ for
both the far-IR detected and far-IR non-detected subsamples. While for
QSO2s, the HR for both the far-IR detected and far-IR non-detected
subsamples are also very similar ($HR\sim -0.12$). 
The large difference in the HR between QSO1s and QSO2s is consistent
with the typical HR values for type I and type II X-ray AGNs, which has been
demonstrated in H07. 
However, we here point out that the average hardness
ratio is insensitive to the presence of far-IR emitting dust. 
The uniform HRs for far-IR detected and far-IR non-detected
  QSO2s might simply be due to the wide redshift distribution from which
  the stacked sources are drawn from, since the
  uncertainties of the measured HRs corresponds to more than 0.3 dex in
  $N_{H}$ ($2-5\times 10^{22}$ cm$^{-2}$, e.g. Fig. 15 in H07) for the redshift range of our sample. 
However, the hardness ratios for the far-IR detected and far-IR
non-detected QSO1s are both consistent with little or no neutral gas
absorption, which cannot explain the large $L_{\rm X}$ difference
between far-IR QSO1s and far-IR non-detected QSO1s.
Therefore, the constant HR regardless of the presence of strong far-IR
emissions might imply that  the lower average $L_X$ and the 
higher $A_{\it V}$ in far-IR detected subsamples are due to gas of large column
density that obscures X-ray photons in both hard-band and
soft-band. We note that large columns of ionized gas residing within
10pc of the vicinity of the X-ray emitting corona might also absorb both the hard-band and
soft-band photons without altering the observed HR. As demonstrated in
various studies of X-ray spectral models \citep[e.g.][]{ceba96}, the
observed hardness ratio is a much weaker function of 
the column density of the ionized warm absorber when compared to a
neutral, cold absorber. This specific type of absorption has been
observed in QSOs hosted by extremely luminous submillimeter galaxies \citep[e.g.][]{page11}, similar to the luminous SF galaxies hosting QSO2s
in this work. However, whether the same type of warm absorber
  exists in our far-IR detected QSOS cannot be confirmed without high-resolution X-ray spectroscopy. 

\begin{figure}
\epsscale{1.2}
\plotone{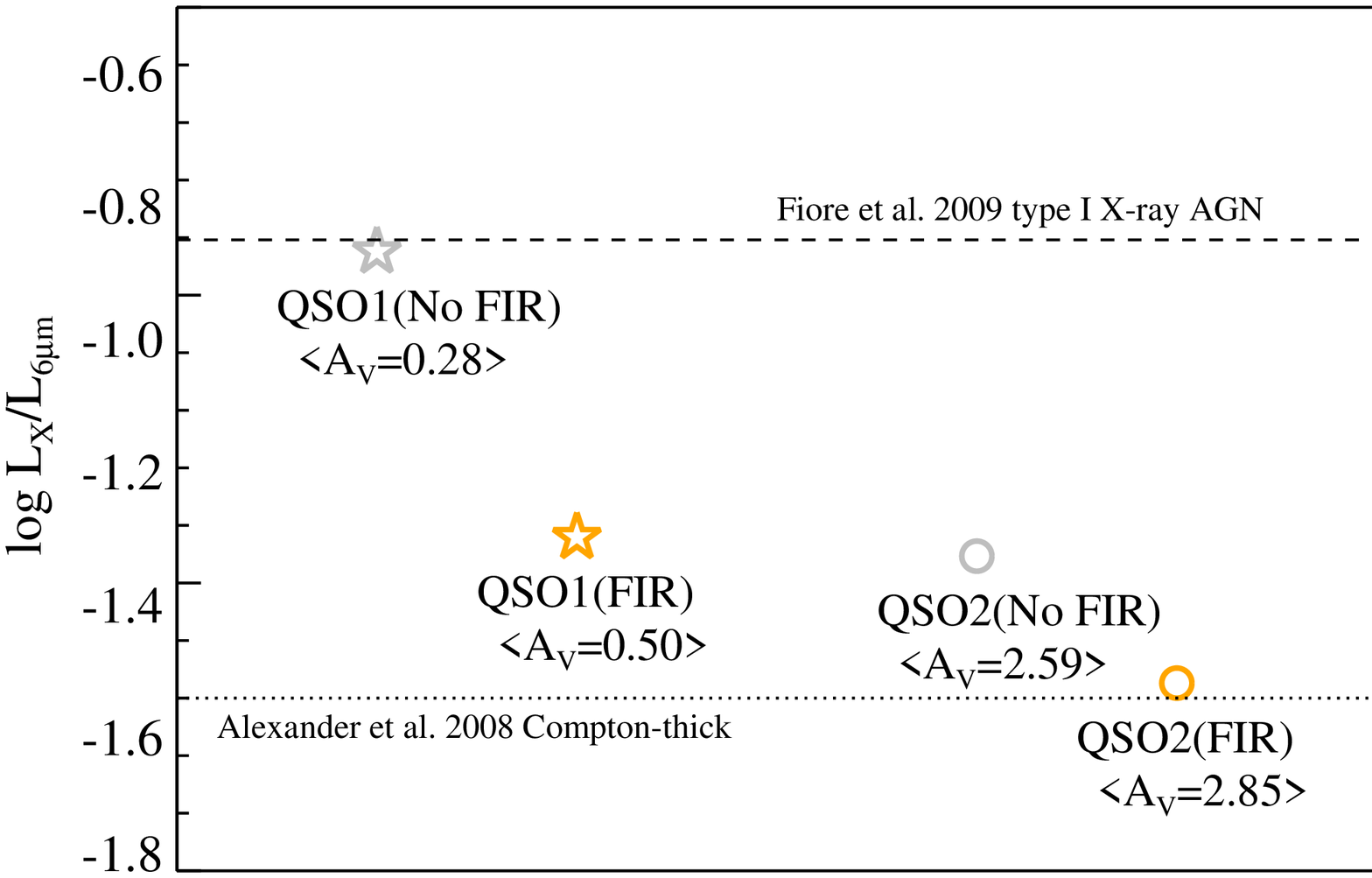}
\caption{A summary of the $L_{X}/L_{6\micron}$ for different QSO
  populations. For comparison, we show the
  $L_{X}/L_{6\micron}$ relationship for type I X-ray AGNs using Eq. 2
  from \cite{fior09obsc} evaluated at $L_{6\micron}=10^{44.9}$\ergs\ as the dashed line. We also show
  the $L_{X}/L_{6\micron}$ relation for local AGNs with Compton-thick
  obscurations \citep{alex08compthick} as the dotted line. We show that the presence of
  far-IR emitting dust attenuates the observed $L_X$ in comparison to
  $L_{6\micron}$ and increases the AGN extinction in the optical
  wavelengths. We note that  far-IR luminous QSO2s might not be detected in
  wide-field X-ray surveys due to the heavy obscuration that can be
  associated with the dust-enshrouded host galaxies. }
\label{fig:lxl6um}
\end{figure}

We summarize the results of our stacking analysis in
Fig.~\ref{fig:lxl6um}, where we show the ratio between
$L_{6\micron}$ and $\langle L_X \rangle$ for different subsamples of
QSOs. Since $L_{\rm X}$  and $L_{\rm MIR}$ are both excellent
tracers of SMBH accretion and $L_{\rm MIR}$ is relatively insensitive to
obscuration, the $L_X$ to $L_{\rm MIR}$ ratio has been used to track down the
elusive population of Compton-thick AGNs and to study the intrinsic
AGN accretion rate \citep[e.g.][]{lutz04irx,alex08compthick,gand09seyfir,ecka10,goul11,rovi13,lans14nustar,ster14nustar}. \par 

In Fig.~\ref{fig:lxl6um}, we show that for both QSO1s and
QSO2s, the existence of a far-IR detection is connected with smaller $L_X/L_{6\micron}$ ratios by at least 0.2 dex when compared to the
far-IR non-detected subsamples. This indirectly supports a link between far-IR
emitting dust and the absorption of X-ray photons. 
For comparison, we also show the luminosity-dependent
$L_{6\micron}-L_X$ relation by \cite{fior09obsc} evaluated
at the average $L_{6\micron}$ of our
sample $L_{6\micron}=10^{44.9}$\ergs\ . We find that the $L_X/L_{6\micron}$ for QSO1s without
direct far-IR detections is similar to the \cite{fior09obsc} relation. 
We note that for local active galaxies, an almost linear
$L_{6\micron}-L_X$ relation has been reported by several studies
\citep[e.g.][]{lutz04irx,maio07,gand09seyfir}, but the inferred $L_X/L_{6\micron}$ value at $L_{6\micron}=10^{44.9}$\ergs\ would be
$-0.37$ if we adopt Eq. 2 from \cite{gand09seyfir}, which is much higher than the results for our far-IR non-detected QSO1s and the
\cite{fior09obsc} relation. The X-ray hardness ratio of the
QSO1 sample is consistent with little or no X-ray
absorption, suggesting that the different $L_X/L_{6\micron}$ ratios for QSOs
and the X-ray selected AGNs \citep{fior09obsc} observed are not simply
due to the difference in gas absorption. The difference in the intrinsic mid-IR
to X-ray spectral shape between local AGNs and quasars is beyond the scope
of this work but will be addressed in a follow-up work by Chen et
al. (2015, in preparation). \par

For AGNs with Compton-thick absorption, the observed $L_X$
can be attenuated by a factor of more than 15 when compared to the
intrinsic $L_{6\micron}-L_X$ relation \citep[e.g.][]{alex08compthick,goul10compthick}. We find that the $L_X/L_{6\micron}$ of
our far-IR detected QSO2s is consistent with the \cite{alex08compthick}
relation derived from attenuating the X-ray luminosities for the
$L_X/L_{6\micron}$ relation of local AGNs \citep{lutz04irx} with Compton-thick material.
However, recent observations have also discovered several
luminous AGNs with extremely weak X-ray emissions even in the
ultra-hard energy bands probed by NuSTAR
\citep{luo13nustar,teng14nustar,lans14nustar,ster14nustar}.
Therefore, it is still unclear whether the low
$L_X/L_{6\micron}$ ratio for the QSO2s is due to Compton-thick
obscuration or intrinsic X-ray weakness. Nonetheless, the low
$L_X/L_{6\micron}$ for far-IR detected QSO2s is another line of evidence indicating that X-ray AGN
selection methods can miss heavily obscured mid-IR QSOs hosted by
galaxies with active SF activity. We caution that all AGN selection criteria suffer from various different selection biases,
especially the incompleteness due to obscuration and the contamination from SF galaxy
interlopers. We reiterate that SF galaxy contamination
with our mid-IR color selection criteria is relatively mild for the shallow
mid-IR flux limits of our sample. Our result demonstrates the advantage of mid-IR AGN selection criteria
in detecting deeply obscured quasars hosted by powerful SF galaxies. \par

\section{Obscured fraction and $L_{\rm IR}^{\rm SF}$}
\begin{figure}[t]
\epsscale{1.2} 
\plotone{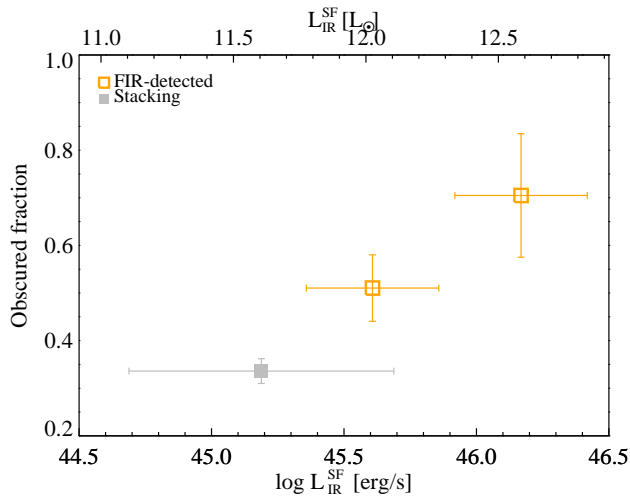}
\caption{The obscured fraction (i.e. the QSO2 fraction) as a
  function of $L_{\rm IR}^{\rm SF}$. We separate far-IR detected QSOs into two
  bins of $L_{\rm IR}^{\rm SF}$ with equal number of sources in each bin, and plot
  their obscured fraction as the orange open squares. For the
  far-IR non-detected QSOs, we show the obscured fraction at the
  $L_{\rm IR}^{\rm SF}$ estimated from stacking analysis as the gray
  filled squares. The horizontal error bars show the inter-percentile
  range in $L_{\rm IR}^{\rm SF}$ containing $80\%$ of the sample. This
  plot shows a monotonic increase of obscured fraction at higher
  $L_{\rm IR}^{\rm SF}$ which connects the AGN obscuration to the host galaxy star formation. }
\label{fig:obsfrac}
\end{figure}

In \S5 and \S6 we have shown that mid-IR selected QSO2s reside in
galaxies with more active SF and that their AGNs are more obscured at both
optical and X-ray wavelengths. While these results support the
connection between AGN obscuration and star formation in the QSO host
galaxies, we next directly test the connection between quasar obscuration and
galaxy SF with the measurement of the ``obscured fraction'' as a
function of $L_{\rm IR}^{\rm SF}$. We show this result in
Fig.~\ref{fig:obsfrac}. 
For the far-IR detected QSOs, where $L_{\rm IR}^{\rm SF}$
can be more accurately estimated, we separate the sources into two bins of
$L_{\rm IR}^{\rm SF}$ with equal number of sources in each bin. For the QSOs with
$L_{\rm IR}^{\rm SF}$ estimated from stacked far-IR fluxes, the $L_{\rm IR}^{\rm SF}$ suffer from larger
uncertainties. Therefore, we do not group these QSOs based on their
$L_{\rm IR}^{\rm SF}$ but only evaluate obscured fractions for the entire
populations. We find that the obscured fraction of the far-IR
detected QSOs is much higher than that of the rest of QSOs, and the
obscured fraction increases monotonically with respect to $L_{\rm
  IR}^{\rm SF}$ as $f_{\rm obs} \propto0.39 \times \log L_{\rm
  IR}^{\rm SF}$.  This agrees with a scenario which connects the obscuration in powerful quasars
to the host galaxy star formation, as the AGN unification model do not
predict an increased obscured fraction for QSOs hosted by galaxies
with higher $L^{\rm SF}_{\rm IR}$. \par

\section{Verification of the difference between QSO1s and QSO2s}
In this work, we have demonstrated that the star formation properties
of QSO1s and QSO2s are different, which implies a link between 
star-forming dust and AGN obscuration in quasars. 
To verify the observed difference between QSO1s and QSO2
is not driven by the uncertainty of photometric redshifts of QSO2s and the presence
of starburst contaminations, we perform several tests on the QSO1
sample and use the spectroscopically confirmed AGNs as a benchmark of
how the various uncertainties might affect the observed properties of
QSO2s.

\subsection{Far-IR detection fraction}
We reiterate that the sources with a dominating starburst
  component have been removed from the final sample studied in this
  analysis based on the SED fits discussed in \S3. Therefore, the higher far-IR detection fraction for QSO2s
is not due to the inclusion of starburst interlopers in the mid-IR AGN
sample. Another possible issue that can drive the higher
far-IR detection fraction for the QSO2s is the \pz\ uncertainties.
We here examine this issue by testing the
null hypothesis that QSO1s and QSO2s have exactly the same far-IR
detection fraction and redshift distribution. \par
From the AGES catalog, we select all of the sources with broad
emission lines (the BLAGNs). In the AGES catalog there are a total of
1619 BLAGNs of which 181 of them have a SPIRE $250\;\micron$ flux
greater than 20 mJy. This far-IR detection fraction is similar to that for the QSO1 sample. To test if the uncertainties in photometric redshifts can boost the
far-IR detection fraction, we randomly scatter the redshifts for all BLAGNs with the conservative uncertainty of
$\sigma_{z} = 0.25(1+z)$ (discussed in \S2.1). In each random realization, we select the
sources with a randomly assigned redshift within $0.7<z<1.8$ and calculate their
far-IR detection fraction. 
We repeat this process 100 times and found
that on average, the random sample would have a far-IR
detection fraction of only $\sim 9\%$. This means that the uncertainty
in the photometric redshift would actually {\it decrease} the observed far-IR detection
fraction if the far-IR luminosity function and the redshift
distribution are the same for QSO1s and QSO2s. This is likely due to
the negative {\it k}-correction at $250\;\micron$ for galaxies with
typical cold dust SEDs \citep[e.g.][]{blai96,negr07}.  The
starburst cold dust emission peaks at rest-frame $\sim 100\;\micron$
which correspond to the SPIRE $250\;\micron$ for an object at
$z=1.5$. Therefore, the starburst galaxies at lower redshifts are not
brighter than our far-IR detected QSOs at observed-frame $250\;\micron$ due to the
negative {\it k}-correction. High-redshift starbursts are also not
likely to be included in our far-IR detected sample as the
$250\;\micron$ filter probes the rapidly dropping
Rayleigh-Jeans tail of the cold dust emission for objects beyond the
redshift range of our sample.

Considering the significant difference of $\sim 17\%$ between the
far-IR detection fractions of QSO1s and QSO2s, we confirm that the
QSO2s are indeed more likely to be hosted by galaxies with strong far-IR emission than QSO1s.

\subsection{Far-IR luminosity}
Since the majority of QSO2s have only \pzs\ and luminosity
measurements are redshift dependent, it is of extreme importance to
confirm that the $0.3$ dex average $\lsf$  difference between QSO1s
and QSO2s is not driven by solely by the uncertainty of \pzs\ . 
Similar to the previous subsection, we estimate the effect of the \pz\ uncertainty by testing the
null-hypothesis that the QSO2s have the same $\lsf\ $ distribution as
the QSO1s. We again randomly assign redshifts to the AGES BLAGNs based
on their {\it spec-z}s and the uncertainty upper limit
$\sigma_z=0.25(1+z)$. For the sources with a randomly assigned
redshift within the $0.7<z<1.8$ range, we perform the SED analysis
(described in \S3) using the assigned redshift to calculate their
average $\lsf\ $. We repeat this process 100 times and found that the average $\lsf\ $ difference between the random samples and
the original QSO1 sample is $0.08\pm0.02$ dex, which is much smaller in
comparison to the observed $0.3$ dex difference between QSO1s and
QSO2s. We also note that the $\sigma_z=0.25(1+z)$ is a very
conservative upper limit so the effects of photometric redshift
uncertainty are almost certainly smaller. The $L_{\rm SF}$ difference
between QSO1s and QSO2s is not driven by the \pz\ uncertainty.

\section{Discussion and Summary}
Recently, a number of studies of X-ray selected AGNs have found that star formation
does not distinguish between AGNs with and without obscuration for
AGNs classified based on either the presence of broad emission lines
\citep{merl13} or the observed gas column density \citep[$N_{\rm
  H}$, e.g.][]{rovi12,rosa12agnsf}.
These conclusions are in direct contrast with our result in
Fig.~\ref{fig:lagn} showing a significant ($> 0.25$ dex) difference
in $L_{\rm IR}^{\rm SF}$ between the QSO1s and the QSO2s. \par

To explain this apparent contradiction, we point out a fundamental difference between these studies and the
present work. \cite{merl13} classifies the X-ray selected AGN into ``type 1'' and ``type 2''
sources based SED fitting or the detection of broad emission lines. 
In our ${\it R}-[4.5]$ classification, all of the QSO1s are
spectroscopically selected type 1 quasars, but it is not clear
how many of our QSO2s would have spectra with narrow emission
lines similar to that of the optical type 2 quasars; since the large amount of dust which blocks most of the AGN optical continuum might very well
block the emission lines from the narrow line region essential to
optical AGN classifications. 
Also, while deep X-ray and optical observations have
proven to be efficient at detecting both obscured and
unobscured AGNs, it is still challenging to include AGNs
obscured by material with column density reaching $N_{H}\sim
5\times10^{24}$ cm$^{-2}$ (i.e. Compton thick) in optical or X-ray
selected samples. 
Recent works on synthetic models of the cosmic X-ray backgrounds
\citep[e.g.][]{gill07cxb,trei09,ball11}  suggest that a significant
fraction of SMBH growth might occur in a heavily obscured, rapidly
accreting AGN phase. As we have shown in Fig.~\ref{fig:lxl6um}, the
far-IR detected QSO2s have a $L_X/L_{6\micron}$ ratio consistent with
AGNs that have heavily absorbed or intrinsically weak X-ray emission. 
Therefore, the different results might simply be due
to mid-IR selected quasars representing different AGN populations than
the optically and X-ray selected AGNs. \par

Indeed, mid-IR selection criteria have been shown to be capable of picking up sources with
their nuclei obscured by Compton-thick material
\citep[e.g.][]{rovi13}, and moderate-luminosity mid-IR AGNs have been found to preferentially
populate the higher-end of AGN Eddington ratio
distributions \citep[e.g.][]{hick09}, and to reside in more blue,
star-forming host galaxies \citep[e.g.][]{hick09,goul13}. In conjunction with
the higher $L_{\rm SF}^{\rm IR}$ and obscuration in our mid-IR quasars, we argue
that part of the enhancement in $L_{\rm IR}^{\rm SF}$ of our sample when compared to X-ray or
optical quasars is due to having different AGN populations based on
different selection criteria and the heavy obscuration which possibly takes place on the scale of the host galaxy. \par

In summary, we have studied mid-IR luminous quasars at
$0.7<z<1.8$ and demonstrated a connection between obscuration in
mid-IR selected QSOs and star formation of their host galaxies.
This connection is supported by the higher AGN obscured fraction in QSO host galaxies
with larger $L_{\rm IR}^{\rm SF}$ and the large differences in the average far-IR luminosities between
QSO1s and QSO2s. We have also shown that both the AGN obscuration at optical
wavelengths and the AGN absorption at X-ray wavelengths can be
connected to the presence of far-IR emitting dust. These results
suggest that the large scale gas and dust are also obscuring the central AGNs in
addition to orientation-based small-scale obscuration, which
is consistent with the scenario in which the rapidly growing SMBHs are
connected to dust-enshrouded starburst galaxies. 

\begin{acknowledgements}
We thank the anonymous referee for a careful reading of the
manuscript and helpful suggestions that strengthened this paper greatly.
We thank our colleagues on the AGES, ISS, SDWFS, NDWFS and the
\xbootes\ teams, and the HerMES team for making the data publicly
available. The first {\em Spitzer} MIPS survey of the \bootes\ region
was obtained using GTO time provided by the {\em Spitzer} Infrared
Spectrograph Team (PI: James Houck) and by M. Rieke.  We thank the
collaborators in that work for access to the $24\;\micron$ catalog
generated from those data by Emeric LeFloc'h. This work was supported
under contract NAS8-03060. R.J.A. was supported by Gemini-CONICYT
grant number 32120009. C.-T.J.C, K.N.H, and R.C.H. were partially supported by the National
Science Foundation through grant No. 1211096. K.N.H. and R.C.H. were
partially supported by NASA through ADAP award NNX12AE38G.
R.C.H. acknowledges support from an Alfred P. Sloan Research
Fellowship and Dartmouth Class of 1962 Faculty Fellowship. C.-T.J.C was supported by a Dartmouth Fellowship and the William H. Neukom 1964 Institute for Computational Science.
\end{acknowledgements}


\end{document}